\documentclass[aps, prb, twocolumn, superscriptaddress,preprintnumbers]{revtex4-2}
\usepackage{bm, amsmath, amsfonts, amssymb, ascmac, mathtools, braket}
\usepackage{multirow}
\usepackage{graphicx}
\usepackage{float, color, xcolor}

\usepackage[whole]{bxcjkjatype} 
\usepackage{subfigure} 
\usepackage{amscd} 

\usepackage{bbm}
\usepackage{tabularx}

\usepackage{comment}

\definecolor{rred}{rgb}{0.8, 0.0, 0.0}
\definecolor{bblue}{rgb}{0.0, 0.0, 0.8}

\usepackage[
pagebackref=false,
colorlinks=true,
linkcolor=bblue,
urlcolor=bblue,
filecolor=black,
citecolor=rred,
pdfstartview=FitV,
pdftitle={},
pdfauthor={},
pdfsubject={},
pdfkeywords={},
pdfpagemode=None,
bookmarksopen=true
]{hyperref}

\newcommand{\ii}{\text{i}}

\begin{document}

\title{Quantum geometry of the non-Hermitian skin effect}

\author{Ken-Ichiro Imura}
\email{imura@iis.u-tokyo.ac.jp}
\affiliation{Institute of Industrial Science, The University of Tokyo
5-1-5 Kashiwanoha, Kashiwa 277-8574, Japan}
\affiliation{RIKEN Center for Emergent Matter Science (CEMS), Wako, Saitama 351-0198, Japan}
\affiliation{Department of Physics, Sophia University, Chiyoda-ku, Tokyo, 102-8554, Japan}

\author{Kohei Kawabata}
\email{kawabata@issp.u-tokyo.ac.jp}
\affiliation{Institute for Solid State Physics, University of Tokyo, Kashiwa, Chiba 277-8581, Japan}

\date{\today}

\begin{abstract}
The non-Hermitian skin effect is nonreciprocity-induced localization phenomena in which a macroscopic number of eigenstates accumulate anomalously at the boundary, 
accompanied by the extreme sensitivity to boundary conditions.
Here, we develop a geometric characterization of the non-Hermitian skin effect.
We demonstrate that the localization length scale associated with the skin effect is encoded in the quantum metric defined solely from right eigenstates, but not in the biorthogonal quantum metric.
Moreover, we show that the quantum metrics can exhibit the power-law divergences at gapless points that depend on the different boundary conditions.
We also reveal that cusps of the generalized Brillouin zone in non-Bloch band theory are signaled by discontinuities in the quantum metrics.
We illustrate these behavior using prototypical non-Hermitian models, such as the Hatano-Nelson model and the non-Hermitian, nonreciprocal Su-Schrieffer-Heeger model.
\end{abstract}

\maketitle

\section{Introduction}

Recent years have seen growing interest in the geometry of quantum wave functions in condensed matter physics~\cite{Yu-25-review, Gao-25-review}. 
While early developments were driven primarily by Berry phases and the associated topological phenomena, it has become increasingly clear that the quantum metric, the symmetric part of the quantum geometric tensor, encodes physically relevant information that is not exhausted by topology alone. 
The quantum metric quantifies the infinitesimal distance between nearby quantum states in parameter space and provides a measure of how rapidly the underlying wave functions change,
which has proved useful in a broad range of contexts. 
For example, it characterizes the spread and localization of Wannier functions~\cite{Marzari-97, Resta-99, Souza-00, Brouder-07}, 
captures the superfluid weight in flat-band superconductors~\cite{Peotta-15, Chen-24}, 
yields bounds for observables~\cite{Onishi-24, Shinada-25},
and provides geometric descriptions of the fractional quantum Hall effect and related correlated phases~\cite{Roy-14, Wang-21, Ledwith-23}. 
These developments have established quantum geometry as a useful framework for describing intrinsic properties of quantum states.

Meanwhile, non-Hermitian systems have emerged as a central theme of contemporary condensed matter physics. 
Over the past several years, topological aspects of non-Hermitian systems have been studied intensively in both theory~\cite{Rudner-09, Sato-11, *Esaki-11, Hu-11, Schomerus-13, Longhi-15, Malzard-15, Leykam-17, Xu-17, Shen-18, *Kozii-17, Takata-18, Gong-18, *Kawabata-19, McDonald-18, Longhi-19, KSUS-19, ZL-19, KBS-19, JYLee-19, Chang-20, Wanjura-20, Bessho-21, Denner-21, *Denner-23JPhysMater, KSR-21, Okugawa-21, Vecsei-21, Shiozaki-21, Franca-22, Nakamura-24, Ma-24, Schindler-23, Nakamura-23, Tanaka-25, WangBenalcazar-25} and experiment~\cite{Poli-15, Zeuner-15, Zhen-15, Weimann-17, Xiao-17, St-Jean-17, Parto-17, Bahari-17, Zhao-18, Zhou-18, Harari-18, *Bandres-18, Cerjan-19, Zhao-19, Wengang-21, WangWangMa-22, Ochkan-24}. 
Non-Hermitian descriptions are physically relevant because they naturally arise as effective descriptions of open classical and quantum systems in realistic settings~\cite{Konotop-review, Christodoulides-review}. 
Among the phenomena unique to non-Hermitian systems, the non-Hermitian skin effect attracts particular attention~\cite{Lee-16, Xiong-18, MartinezAlvarez-18, YW-18-SSH, *YSW-18-Chern, Kunst-18, KSU-18, Lee-Thomale-19, Liu-19, Lee-Li-Gong-19, Herviou-19, Zirnstein-19, Borgnia-19, Yokomizo-19, Imura-19, Zhang-20, OKSS-20, *KOS-20, Yang-20, Yoshida-20, Yi-Yang-20, Xue-20, Okugawa-20, KSS-20, Fu-21, Zhang-22, Sun-21, Wang-24, Nakai-24, Zhang-25}. 
The non-Hermitian skin effect is a nonreciprocity-induced localization phenomenon in which a macroscopic number of eigenstates accumulate anomalously at the boundary, accompanied by an extreme sensitivity to boundary conditions. 
It has been observed experimentally in a wide variety of classical and quantum platforms~\cite{Brandenbourger-19-skin-exp, *Ghatak-19-skin-exp, Helbig-19-skin-exp, *Hofmann-19-skin-exp, Xiao-19-skin-exp, Weidemann-20-skin-exp, Gou-20, *Liang-22, Palacios-21, Zhang-21, Wang-23mech, Liu-24, Zhao-25, Shen-25, Wu-24}.
Its theoretical understanding has been developed through non-Bloch band theory, in which the ordinary Brillouin zone is replaced by a generalized Brillouin zone with complex-valued momentum~\cite{YW-18-SSH, Yokomizo-19}. 
This framework has clarified how the bulk properties of non-Hermitian systems are described under open boundary conditions, and how the bulk-boundary correspondence is reformulated in non-Hermitian systems.

Despite this progress, the geometric structure of non-Hermitian systems remains much less developed. 
Whereas non-Hermitian topological phases have been investigated extensively, more general geometric characteristics of non-Hermitian systems have only recently begun to receive attention~\cite{Ye-24, Hu-25, Behrends-25, *Matraszek-25, Chen-26, Montag-26}. 
This point is especially significant for the non-Hermitian skin effect. 
The skin effect is a phenomenon of eigenstates that concerns their spatial accumulation, sensitivity to boundaries, and the deformation of the Brillouin zone.
Quantum geometry, on the other hand, is precisely a framework for quantifying how eigenstates vary in parameter space. 
It is therefore natural that these two subjects should be closely related.
Nevertheless, such a relation has not been clarified. 
Moreover, in non-Hermitian systems, since right and left eigenstates are generally distinct, quantum geometry is no longer unique, and different definitions do not necessarily encode the same physical content. 
Establishing which notion of quantum metric captures the skin effect, and which geometric features of non-Hermitian band structures can be diagnosed in this manner, is therefore a fundamental problem.

In this work, we develop a geometric characterization of the non-Hermitian skin effect. 
Our starting point is the observation that, in Hermitian systems, the quantum metric provides the localization scale of wave functions. 
We show that an analogous statement holds for the skin effect, but with an important modification that has no counterpart in Hermitian band theory. 
Specifically, we demonstrate that the localization length associated with skin modes is encoded in the quantum metric defined solely from right eigenstates, whereas it is not encoded in the biorthogonal quantum metric defined from both right and left eigenstates. 
This distinction reveals that the difference between right and left eigenstates directly affects which geometric quantity is physically relevant to localization. 
We illustrate this result with the Hatano-Nelson model~\cite{Hatano-Nelson-96, Hatano-Nelson-97}, in which the localization aspect of the skin effect is most transparent.

Furthermore, we show that the quantum metrics capture gap-closing points appropriate to the boundary conditions imposed, thereby reflecting the boundary-sensitive nature of non-Hermitian systems. 
In addition, we demonstrate that the cusp singularities of the generalized Brillouin zone, which are characteristic features of non-Bloch band theory, are signaled by discontinuous behavior of the quantum geometry.
We exemplify these results with another prototypical non-Hermitian model hosting the skin effect, the non-Hermitian, nonreciprocal Su-Schrieffer-Heeger model and its generalizations~\cite{Lee-16, YW-18-SSH, Kunst-18, Yokomizo-19}, for which internal band geometry and the generalized Brillouin zone play essential roles.
Although the cusp singularities of the generalized Brillouin zone were discussed in previous studies~\cite{YW-18-SSH, Yokomizo-19},
it has remained largely unclear which physical quantities directly encode them.
We show that such nonanalytic structures leave a gauge-invariant imprint on quantum geometry.

It is also notable that the quantum geometry of the non-Hermitian Su-Schrieffer-Heeger model was also investigated under the periodic boundary conditions~\cite{Ye-24}.
In this work, we rather focus on the open boundary conditions and clarify how the non-Hermitian skin effect leaves its imprint on the quantum geometry.
Our work identifies quantum geometry as a useful framework for characterizing both the localization physics and the nonanalytic geometric structures induced by the non-Hermitian skin effect.
It should also be noted that the quantum geometric tensor measures the gauge-independent part of the parameter dependence of eigenstates and is known to enter a variety of physical responses in Hermitian systems~\cite{Yu-25-review, Gao-25-review}.
Analogous roles are therefore expected in non-Hermitian systems.

The rest of this work is organized as follows. 
In Sec.~\ref{sec: QG}, we introduce the quantum metrics defined from right eigenstates and from biorthogonal pairs of right and left eigenstates.
In Sec.~\ref{sec: Hatano-Nelson}, we study the Hatano-Nelson model and show that the quantum metric defined solely from right eigenstates captures the localization length associated with the non-Hermitian skin effect, whereas the biorthogonal metric does not. 
In Sec.~\ref{sec: NH-SSH}, we turn to the non-Hermitian, nonreciprocal Su-Schrieffer-Heeger model and analyze the internal contributions to the quantum metrics, with particular emphasis on their behavior around boundary-condition-dependent gap-closing points. 
In Sec.~\ref{sec: non-Bloch}, we extend the analysis to more general non-Bloch band structures and show that cusp singularities of the generalized Brillouin zone are reflected in discontinuous behavior of the quantum geometry.
Finally, Sec.~\ref{sec: conclusion} is devoted to the summary and discussion.

\section{Quantum geometry in non-Hermitian systems}
    \label{sec: QG}

Quantum geometry characterizes how a quantum state varies in parameter space after removing the trivial dependence on its overall phase or normalization. 
It provides a natural language for quantifying the sensitivity of eigenstates to external parameters and will serve as the basic tool in the following discussion.

\subsection{Quantum geometric tensor in Hermitian systems}

We begin with a brief review of the Hermitian case~\cite{Yu-25-review, Gao-25-review}. 
Let $H(\bm{\lambda})$ be a Hermitian Hamiltonian that depends smoothly on a set of parameters $\bm{\lambda}=\left(\lambda_\mu\right)$, and let $\ket{u_n(\bm{\lambda})}$ be a normalized eigenstate of an isolated band:
\begin{equation}
H(\bm{\lambda}) \ket{u_n(\bm{\lambda})}
=
E_n(\bm{\lambda}) \ket{u_n(\bm{\lambda})},
\quad
\braket{u_n(\bm{\lambda})|u_n(\bm{\lambda})}=1.
\end{equation}
By introducing $\partial_\mu \coloneqq \partial/\partial\lambda_\mu$, the quantum geometric tensor (or equivalently, Fubini-Study metric) is defined as
\begin{equation}
\chi_{\mu\nu}^{\mathrm{H}}
\coloneqq
\braket{\partial_\mu u_n|
\left(1-\ket{u_n}\bra{u_n}\right)
|\partial_\nu u_n}.
\label{eq:qgt_hermitian}
\end{equation}
Its real symmetric part gives the quantum metric tensor,
\begin{equation}
g_{\mu\nu}^{\mathrm{H}}
\coloneqq
\mathrm{Re}\,\chi_{\mu\nu}^{\mathrm{H}},
\label{eq:qmetric_hermitian}
\end{equation}
while its imaginary antisymmetric part yields the Berry curvature,
\begin{equation}
F_{\mu\nu}^{\mathrm{H}}
\coloneqq
-2\,\mathrm{Im}\,\chi_{\mu\nu}^{\mathrm{H}}.
\label{eq:berry_hermitian}
\end{equation}
The quantum geometric tensor is invariant under a local phase transformation
$\ket{u_n(\bm{\lambda})}\to e^{\ii\varphi(\bm{\lambda})}\ket{u_n(\bm{\lambda})}$,
and measures the infinitesimal distance between nearby rays in Hilbert space.
In particular, for a one-dimensional parameter $\lambda$, the diagonal component $g_{\lambda\lambda}^{\mathrm{H}}$ quantifies how rapidly the state changes as $\lambda$ is varied.

\subsection{Two quantum geometric tensors in non-Hermitian systems}

For a non-Hermitian Hamiltonian, right and left eigenstates are generally distinct~\cite{Brody-14}. 
They are denoted by
\begin{equation}
H(\bm{\lambda}) \ket{u_n^{\mathrm R}(\bm{\lambda})}
=
E_n(\bm{\lambda}) \ket{u_n^{\mathrm R}(\bm{\lambda})},
    \label{eq:right_eigen_general}
\end{equation}
and
\begin{equation}
\langle\!\langle u_n^{\mathrm L}(\bm{\lambda}) | H(\bm{\lambda})
=
E_n(\bm{\lambda}) \langle\!\langle u_n^{\mathrm L}(\bm{\lambda}) |.
    \label{eq:left_eigen_general}
\end{equation}
Because of this distinction, extension of Eq.~\eqref{eq:qgt_hermitian} to non-Hermitian systems is not unique. 
Two natural definitions are particularly useful for the present work, as introduced below.

The first one is defined solely from the right eigenstates. 
Assuming the normalization
\begin{equation}
\braket{u_n^{\mathrm R}(\bm{\lambda})|u_n^{\mathrm R}(\bm{\lambda})}=1,
\label{eq:right_normalization_general}
\end{equation}
we introduce
\begin{equation}
\chi_{\mu\nu}^{\mathrm{RR}}
\coloneqq
\braket{\partial_\mu u_n^{\mathrm R}|
\left(1-\ket{u_n^{\mathrm R}}\bra{u_n^{\mathrm R}}\right)
|\partial_\nu u_n^{\mathrm R}}.
\label{eq:qmetric_rr_general}
\end{equation}
This quantity is the direct analog of the Hermitian quantum geometric tensor constructed with the ordinary inner product.
For $\mu=\nu$, it is real and nonnegative.
Moreover, it is invariant under a local phase transformation of the normalized right eigenstate.

The second one is the biorthogonal quantum geometric tensor, for which we impose
\begin{equation}
\langle\!\langle u_n^{\mathrm L}(\bm{\lambda}) | u_n^{\mathrm R}(\bm{\lambda}) \rangle = 1,
\label{eq:biorthonormality_general}
\end{equation}
and define
\begin{equation}
\chi_{\mu\nu}^{\mathrm{LR}}
\coloneqq
\langle\!\langle \partial_\mu u_n^{\mathrm L} |
\left(1-\ket{u_n^{\mathrm R}} \langle\!\langle u_n^{\mathrm L} | \right)
\ket{\partial_\nu u_n^{\mathrm R}}.
\label{eq:qmetric_lr_general}
\end{equation}
Whereas the diagonal components $\chi_{\mu\mu}^{\mathrm{RR}}$ of the right-right geometric tensor are always real and nonnegative, 
the diagonal components $\chi_{\mu\mu}^{\mathrm{LR}}$ of the biorthogonal tensor are generally complex.
Its gauge ambiguity is broader: 
it is invariant under the $\mathrm{GL} \left( 1, \mathbb{C} \right)$ transformation
\begin{equation}
\ket{u_n^{\mathrm R}} \to e^{f(\bm{\lambda})}\ket{u_n^{\mathrm R}},
\qquad
\langle\!\langle u_n^{\mathrm L} | \to e^{-f(\bm{\lambda})} \langle\!\langle u_n^{\mathrm L} |,
    \label{eq:gl_gauge_general}
\end{equation}
with an arbitrary complex function $f(\bm{\lambda}) \in \mathbb{C}$, 
which preserves the biorthonormality condition in Eq.~\eqref{eq:biorthonormality_general}.
Although we refer to $\chi_{\mu\nu}^{\mathrm{LR}}$ as a biorthogonal quantum geometric tensor following the terminology in the literature, it is not necessarily interpreted as a positive-definite metric that measures the distance between eigenstates. 
Rather, $\chi_{\mu\nu}^{\mathrm{LR}}$ should be understood as a complex, $\mathrm{GL} \left( 1, \mathbb{C} \right)$-gauge-invariant biorthogonal quantum geometric tensor, or equivalently as a biorthogonal geometric susceptibility, that characterizes the parameter dependence of the biorthogonal eigenstates after the gauge component is removed.

In the Hermitian limit, where left and right eigenstates coincide, Eqs.~\eqref{eq:qmetric_rr_general} and \eqref{eq:qmetric_lr_general} reduce to the same object in Eq.~\eqref{eq:qgt_hermitian}. 
Away from the Hermitian limit, however, they characterize different aspects of non-Hermitian quantum geometry. 
One of the central purposes of this work is to compare these two quantities and clarify which one captures the non-Hermitian skin effect.
Notably, once the normalization condition in Eq.~\eqref{eq:right_normalization_general} is imposed, the right eigenstate $\ket{u_n^{\mathrm R}}$ is fixed up to an irrelevant phase.
Consequently, the corresponding right-right geometric tensor $\chi_{\mu\nu}^{\mathrm{RR}}$ in Eq.~\eqref{eq:qmetric_rr_general} is independent of the biorthogonal normalization condition in Eq.~\eqref{eq:biorthonormality_general}.
On the other hand, the biorthogonal tensor $\chi_{\mu\nu}^{\mathrm{LR}}$ in Eq.~\eqref{eq:qmetric_lr_general} does not depend on Eq.~\eqref{eq:right_normalization_general} but solely on Eq.~\eqref{eq:biorthonormality_general}.
While we may also introduce a quantum geometric tensor solely in terms of left eigenstates, it should be essentially equivalent to the right-right geometric tensor $\chi^{\rm RR}_{\mu\nu}$.

In the following, we mainly focus on diagonal components associated with a single parameter. 
For brevity, we write
\begin{equation}
\chi^{\mathrm{RR}}(\lambda)
\coloneqq
\chi_{\lambda\lambda}^{\mathrm{RR}},
\qquad
\chi^{\mathrm{LR}}(\lambda)
\coloneqq
\chi_{\lambda\lambda}^{\mathrm{LR}}.
\label{eq:diagonal_metrics_general}
\end{equation}
The relevant parameter will be the crystal momentum under the periodic boundary conditions, or the coordinate that parametrizes the generalized Brillouin zone under the open boundary conditions.

\subsection{Plane-wave and internal contributions}
    \label{sec: plane-wave vs internal}

For lattice systems, it is useful to distinguish between the geometry associated with the spatial profile of a wave function and that associated with internal degrees of freedom such as sublattice structure. 
Suppose that the full right eigenstate factorizes as
\begin{equation}
\ket{\psi_n^{\mathrm R}(\bm{\lambda})}
=
\ket{\phi^{\mathrm R}(\bm{\lambda})}
\otimes
\ket{u_n^{\mathrm R}(\bm{\lambda})},
\label{eq:factorized_right_state_general}
\end{equation}
where $\ket{\phi^{\mathrm R}}$ represents the plane-wave part and $\ket{u_n^{\mathrm R}}$ the internal part. 
Similarly, let the left eigenstate factorize as
\begin{equation}
\langle\!\langle \psi_n^{\mathrm L}(\bm{\lambda}) |
=
\langle\!\langle \phi^{\mathrm L}(\bm{\lambda}) |
\otimes
\langle\!\langle u_n^{\mathrm L}(\bm{\lambda}) |,
    \label{eq:factorized_left_state_general}
\end{equation}
with
\begin{equation}
\braket{\phi^{\mathrm R}|\phi^{\mathrm R}}=\langle\!\langle \phi^{\mathrm L} |\phi^{\mathrm R}\rangle=1.
    \label{eq:plane_wave_normalization_general}
\end{equation}
Then, the quantum geometric tensors decompose additively into plane-wave and internal contributions:
\begin{align}
\chi_{\mu\nu}^{\mathrm{RR}}
&=
\chi_{\mu\nu}^{\mathrm{RR,\,pw}}
+
\chi_{\mu\nu}^{\mathrm{RR,\,int}},
\label{eq:decomposition_rr_general} \\
\chi_{\mu\nu}^{\mathrm{LR}}
&=
\chi_{\mu\nu}^{\mathrm{LR,\,pw}}
+
\chi_{\mu\nu}^{\mathrm{LR,\,int}},
\label{eq:decomposition_lr_general}
\end{align}
with
\begin{align}
\chi_{\mu\nu}^{\mathrm{RR,\,pw}}
&\coloneqq
\braket{\partial_\mu \phi^{\mathrm R}|
\left(1-\ket{\phi^{\mathrm R}}\bra{\phi^{\mathrm R}}\right)
|\partial_\nu \phi^{\mathrm R}},
    \label{eq:pw_rr_general} \\
\chi_{\mu\nu}^{\mathrm{RR,\,int}}
&\coloneqq
\braket{\partial_\mu u_n^{\mathrm R}|
\left(1-\ket{u_n^{\mathrm R}}\bra{u_n^{\mathrm R}}\right)
|\partial_\nu u_n^{\mathrm R}},
    \label{eq:int_rr_general} \\
\chi_{\mu\nu}^{\mathrm{LR,\,pw}}
&\coloneqq
\langle\!\langle \partial_\mu \phi^{\mathrm L} |
\left(1-\ket{\phi^{\mathrm R}} \langle\!\langle \phi^{\mathrm L} | \right)
\ket{\partial_\nu \phi^{\mathrm R}}, 
    \label{eq:pw_lr_general} \\
\chi_{\mu\nu}^{\mathrm{LR,\,int}}
&\coloneqq
\langle\!\langle \partial_\mu u_n^{\mathrm L} |
\left(1-\ket{u_n^{\mathrm R}}\langle\!\langle u_n^{\mathrm L} | \right)
\ket{\partial_\nu u_n^{\mathrm R}}.
    \label{eq:int_lr_general}
\end{align}

This distinction is especially relevant to the present problem. 
The plane-wave contribution is already nontrivial even in models without internal degrees of freedom, and it is directly sensitive to real-space localization. 
For an extended plane wave under periodic boundary conditions, it typically grows with the system size.
However, as shown in Sec.~\ref{sec: Hatano-Nelson}, under the open boundary conditions in non-Hermitian systems, such a plane wave can become localized as a consequence of the non-Hermitian skin effect, and the plane-wave contribution to the quantum geometry can then be governed by the corresponding localization length scale.
Specifically, in Sec.~\ref{sec: Hatano-Nelson}, we analyze a model without internal degrees of freedom and show that the metric defined from right eigenstates detects the localization length associated with the non-Hermitian skin effect.
By contrast, the internal contribution becomes essential in multiband models, where the quantum geometry probes the parameter dependence of the internal eigenstate and can become singular at band-touching points.
As we show in Secs.~\ref{sec: NH-SSH} and \ref{sec: non-Bloch}, in the generalized Brillouin-zone formulation, it can also diagnose nonanalytic structures such as cusps.
This framework allows us to compare the metric defined solely from right eigenstates with the biorthogonal one on equal footing throughout this work.

\subsection{
Time-reversal symmetry}
    \label{sec: symmetry}

Symmetry of the non-Hermitian Hamiltonian $H \left( \bm{\lambda} \right)$ constrains its eigenstates and associated quantum metrics.
Specifically, suppose that $H \left( \bm{\lambda} \right)$ respects
\begin{equation}
    \mathcal{T} H^{*} \left( \bm{\lambda} \right) \mathcal{T}^{-1} = H \left( - \bm{\lambda} \right), \quad \mathcal{T}\mathcal{T}^* = +1,
        \label{eq: TRS}
\end{equation}
with a unitary matrix $\mathcal{T}$.
This symmetry can be interpreted as time-reversal symmetry when $\bm{\lambda}$ represents momenta, or its generalization in non-Bloch band theory.
Notably, the models studied in this work---the Hatano-Nelson model and the non-Hermitian Su-Schrieffer-Heeger model---respect time-reversal symmetry in Eq.~\eqref{eq: TRS}.
As a result of this symmetry, a right eigenstate $\ket{u_n^{\rm R} \left( \bm{\lambda} \right)}$ in Eq.~\eqref{eq:right_eigen_general} satisfies
\begin{equation}
    H \left( - \bm{\lambda} \right) \mathcal{T} \ket{u_n^{\rm R} \left( \bm{\lambda} \right)}^* = ( E_n \left( \bm{\lambda} \right) )^* \mathcal{T} \ket{u_n^{\rm R} \left( \bm{\lambda} \right)}^*,
\end{equation}
showing that $\mathcal{T} \ket{u_n^{\rm R} \left( \bm{\lambda} \right)}^*$ is a right eigenstate of $H \left( - \bm{\lambda} \right)$ with complex eigenenergy $( E_n \left( \bm{\lambda} \right) )^*$.
Accordingly, we have
\begin{equation}
    \ket{u_n^{\rm R} \left( - \bm{\lambda} \right)} = \mathcal{T} \ket{u_n^{\rm R} \left( \bm{\lambda} \right)}^*
\end{equation}
under the appropriate choice of the gauge.
Similarly, a left eigenstate $| u_n^{\rm L} \left( \bm{\lambda} \right) \rangle\!\rangle$ in Eq.~\eqref{eq:left_eigen_general} satisfies
\begin{equation}
    H^{\dag} \left( - \bm{\lambda} \right) \mathcal{T} | u_n^{\rm L} \left( \bm{\lambda} \right) \rangle\!\rangle^* = E_n \left( \bm{\lambda} \right) \mathcal{T} | u_n^{\rm L} \left( \bm{\lambda} \right) \rangle\!\rangle^*,
\end{equation}
leading to
\begin{equation}
    | u_n^{\rm L} \left( - \bm{\lambda} \right) \rangle\!\rangle = \mathcal{T} | u_n^{\rm L} \left( \bm{\lambda} \right) \rangle\!\rangle^*.
\end{equation}
Consequently, the quantum geometric tensors defined in Eqs.~\eqref{eq:qmetric_rr_general} and \eqref{eq:qmetric_lr_general} satisfy
\begin{align}
    ( \chi_{\mu\nu}^{\rm RR} \left( \bm{\lambda} \right) )^* = \chi_{\mu\nu}^{\rm RR} \left( -\bm{\lambda} \right), \\
    ( \chi_{\mu\nu}^{\rm LR} \left( \bm{\lambda} \right) )^* = \chi_{\mu\nu}^{\rm LR} \left( -\bm{\lambda} \right).
\end{align}
It follows that the real parts of the quantum metrics are even functions of $\bm{\lambda}$, whereas the imaginary parts are odd.

\section{Hatano-Nelson model}
    \label{sec: Hatano-Nelson}

As a paradigmatic example that exhibits the non-Hermitian skin effect, we consider the Hatano-Nelson model~\cite{Hatano-Nelson-96, *Hatano-Nelson-97}. 
We show that the quantum geometric tensor defined solely from right eigenstates detects the localization length scale of the skin effect, whereas the biorthogonal quantum geometric tensor constructed from both right and left eigenstates does not.

We study the single-particle Hamiltonian $H$ on a chain of length $L$. 
Under the periodic boundary conditions, we take
\begin{equation}
H_{\mathrm{PBC}}
=
\sum_{n=1}^{L}
\left(
\frac{J+\gamma}{2}\,\lvert n+1 \rangle \langle n \rvert
+
\frac{J-\gamma}{2}\,\lvert n \rangle \langle n+1 \rvert
\right),
\end{equation}
with $\lvert L+1 \rangle \coloneqq \lvert 1 \rangle$.
Here, $\ket{n}$'s ($n=1, 2, \cdots, L$) denote the single-particle sites.
Under the open boundary conditions, we instead consider
\begin{equation}
H_{\mathrm{OBC}}
=
\sum_{n=1}^{L-1}
\left(
\frac{J+\gamma}{2}\,\lvert n+1 \rangle \langle n \rvert
+
\frac{J-\gamma}{2}\,\lvert n \rangle \langle n+1 \rvert \right).
    \label{eq: Hatano-Nelson OBC}
\end{equation}
Here, $J \in \mathbb{R}$ denotes the amplitude of the reciprocal hopping while $\gamma \in \mathbb{R}$ characterizes the amplitude of the nonreciprocal hopping.

\subsection{Periodic boundary conditions}

Under the periodic boundary conditions, both right and left eigenstates are extended Bloch waves,
\begin{equation}
\lvert u(k) \rangle
=
\lvert u(k) \rangle\!\rangle
=
\frac{1}{\sqrt{L}}
\sum_{n=1}^{L}
e^{\ii k n}\,
\lvert n \rangle,
\quad
k \in [0,2\pi).
\end{equation}
Since their derivative with respect to $k$ is
\begin{equation}
\partial_k \lvert u(k) \rangle
=
\frac{\ii}{\sqrt{L}}
\sum_{n=1}^{L}
n e^{\ii k n}\,
\lvert n \rangle ,
\end{equation}
it follows that
\begin{align}
\langle u(k) \rvert \partial_k u(k) \rangle
&=
\frac{\ii}{L}\sum_{n=1}^{L} n
\eqqcolon \ii \langle n \rangle,
\\
\langle \partial_k u(k) \rvert \partial_k u(k) \rangle
&=
\frac{1}{L}\sum_{n=1}^{L} n^2
\eqqcolon \langle n^2 \rangle .
\end{align}
The quantum metric then reads
\begin{align}
\chi(k)
&=
\langle \partial_k u(k) \rvert
\bigl( 1 - \lvert u(k) \rangle \langle u(k) \rvert \bigr)
\lvert \partial_k u(k) \rangle
\nonumber\\
&=
\langle n^2 \rangle - \langle n \rangle^2
\nonumber\\
&= \frac{L^2-1}{12}.
\end{align}
Thus, the quantum metric $\chi$ gives the variance of the center of the Bloch wave and scales as $\chi \propto L^2$ for large $L$, which reflects the delocalization of the eigenstate over the entire system.

\subsection{Open boundary conditions}

Under the open boundary conditions, the nonreciprocal hopping can be effectively removed by a nonunitary similarity transformation (i.e., imaginary gauge transformation in Ref.~\cite{Hatano-Nelson-96, *Hatano-Nelson-97}). 
Introducing
\begin{equation}
V
\coloneqq
\operatorname{diag} \left( b,b^2,\dots,b^L \right),
\qquad
b
\coloneqq
\sqrt{\frac{J+\gamma}{J-\gamma}},
\end{equation}
we have
\begin{equation}
V^{-1} H_{\mathrm{OBC}} V
=
\frac{\sqrt{J^2-\gamma^2}}{2} \sum_{n=1}^{L-1}
\left( 
\lvert n+1 \rangle \langle n \rvert
+
\lvert n \rangle \langle n+1 \rvert
\right).
\end{equation}
This transformed Hamiltonian is Hermitian, and its eigenstates are extended. Consequently, the right and left eigenstates of the original non-Hermitian Hamiltonian, which are obtained by applying $V^{-1}$ and $V$ to the extended eigenstates, acquire exponential factors $b^{\pm n}$. 
It is therefore natural to introduce the complex momentum
\begin{equation}
k + \ii g,
\qquad
g \coloneqq \log b = \frac{1}{2}\log\frac{J+\gamma}{J-\gamma},
\end{equation}
where $g$ characterizes the skin localization and $1/\lvert g \rvert$ plays the role of the localization length.

Accordingly, the right and left eigenstates are obtained as
\begin{align}
\lvert u^{\rm R}(k,g) \rangle
&\simeq
\frac{1}{\sqrt{L}}
\sum_{n=1}^{L}
e^{(\ii k + g)n}\,
\lvert n \rangle,
    \label{eq: ur}
\\
\lvert u^{\rm L}(k,g) \rangle\!\rangle
&\simeq
\frac{1}{\sqrt{L}}
\sum_{n=1}^{L}
e^{(\ii k - g)n}\,
\lvert n \rangle.
    \label{eq: ul}
\end{align}
These states satisfy the biorthogonal normalization condition
\begin{equation}
\langle\!\langle u^{\rm L}(k,g) \vert u^{\rm R}(k,g) \rangle = 1,
\end{equation}
although neither the right nor the left eigenstate is individually normalized in the usual sense (see below).
Notably, the states $\lvert u^{\rm R}(k,g) \rangle$ and $\lvert u^{\rm L}(k,g) \rangle\!\rangle$ in Eqs.~\eqref{eq: ur} and \eqref{eq: ul} can be regarded as being parameterized by $k$ and $g$.
We thus evaluate the quantum geometric tensor in the $\left( k, g \right)$ parameter space, as detailed below.
It should also be noted that Eqs.~\eqref{eq: ur} and \eqref{eq: ul} do not yield exact eigenstates of Eq.~\eqref{eq: Hatano-Nelson OBC}.
Nevertheless, they capture the leading contributions of the skin states, 
in the same manner as the non-Bloch band theory~\cite{YW-18-SSH, Yokomizo-19}; 
see Sec.~\ref{sec: non-Bloch} for further details.

\subsubsection{Quantum geometric tensor from right eigenstates}

We first consider the quantum geometric tensor defined solely from the right eigenstate:
\begin{equation}
\chi_{\mu\nu}^{\mathrm{RR}}
\coloneqq
\langle \partial_\mu u^{\rm R} \rvert
\bigl(1-\lvert u^{\rm R}\rangle\langle u^{\rm R}\rvert\bigr)
\lvert \partial_\nu u^{\rm R} \rangle,
\quad
\mu,\nu \in \{k,g\}.
\label{eq:gRR_HN}
\end{equation}
For this purpose, we normalize the right eigenstate by itself (i.e., $\braket{u^{\rm R} | u^{\rm R}} = 1$):
\begin{equation}
\lvert u^{\rm R} \rangle
=
\frac{1}{\sqrt{Z(g)}}
\sum_{n=1}^{L}
e^{(\ii k + g)n}\,
\lvert n \rangle,
\end{equation}
with
\begin{equation}
Z(g)
\coloneqq
\sum_{n=1}^{L} e^{2gn}
=
\frac{e^{2g}\bigl(e^{2gL}-1\bigr)}{e^{2g}-1}.
\end{equation}
We then obtain
\begin{align}
\partial_k \lvert u^{\rm R} \rangle
&=
\frac{\ii}{\sqrt{Z(g)}}
\sum_{n=1}^{L}
n\,e^{(\ii k + g)n}\,
\lvert n \rangle ,
\\
\partial_g \lvert u^{\rm R} \rangle
&=
\frac{1}{\sqrt{Z(g)}}
\sum_{n=1}^{L}
n\,e^{(\ii k + g)n}\,
\lvert n \rangle \nonumber \\
&\quad -
\frac{1}{Z(g)^{3/2}}
\left(
\sum_{n=1}^{L} n\,e^{2gn}
\right)
\left(
\sum_{n=1}^{L} e^{(\ii k+g)n}\,\lvert n\rangle
\right)
\nonumber\\
&=
-\ii\,\partial_k \lvert u^{\rm R} \rangle
-
\langle n \rangle_g \lvert u^{\rm R} \rangle ,
\end{align}
where we introduce
\begin{equation}
\langle f(n) \rangle_g
\coloneqq
\frac{1}{Z(g)}
\sum_{n=1}^{L}
e^{2gn} f(n).
    \label{eq: f-distribution}
\end{equation}
In particular, we have
\begin{align}
\langle n \rangle_g
&=
-\frac{1}{e^{2g}-1}
+
\frac{L e^{2gL}}{e^{2gL}-1},
\\
\langle n^2 \rangle_g
&=
-\frac{1}{e^{2g}-1}
+
\frac{2}{(e^{2g}-1)^2}
\nonumber \\
&\qquad -
\frac{2L e^{2gL}}{(e^{2g}-1)(e^{2gL}-1)}
+
\frac{L^2 e^{2gL}}{e^{2gL}-1}.
\end{align}
Using these expressions, we find
\begin{align}
\chi_{kk}^{\mathrm{RR}}
&=
\langle n^2 \rangle_g - \langle n \rangle_g^2 \nonumber \\
&=
\frac{e^{2g}}{(e^{2g}-1)^2}
-
\frac{L^2 e^{2gL}}{(e^{2gL}-1)^2} \nonumber \\
&\xrightarrow[L\to\infty]{}
\frac{1}{4\sinh^2 g},
\end{align}
and
\begin{equation}
\chi_{gg}^{\mathrm{RR}}
= \ii \chi_{kg}^{\mathrm{RR}}
= -\ii \chi_{gk}^{\mathrm{RR}}
= \chi_{kk}^{\mathrm{RR}}.
\end{equation}
Hence, the quantum geometric tensor takes the compact form,
\begin{equation}
\chi^{\mathrm{RR}}
=
\chi_{kk}^{\mathrm{RR}}
\begin{pmatrix}
1 & -\ii \\
\ii & 1
\end{pmatrix},
\qquad
\chi_{kk}^{\mathrm{RR}}
\xrightarrow[L\to\infty]{}
\frac{1}{4\sinh^2 g}.
\end{equation}
As in the case of the periodic boundary conditions, the quantum metric $\chi^{\mathrm{RR}}$ describes the variance of the center of the right eigenstate with respect to the deformed distribution induced by the skin effect in Eq.~\eqref{eq: f-distribution}.
Importantly, the localization length scale $g$ due to the skin effect explicitly appears in $\chi^{\mathrm{RR}}$. 
In this sense, the quantum metric defined solely from right eigenstates directly captures the non-Hermitian skin effect.

The corresponding Berry curvature in the complex-momentum space $\left( k, g \right)$ is nonzero:
\begin{equation}
F_{kg}^{\mathrm{RR}}
=
-2\,\operatorname{Im}\chi_{kg}^{\mathrm{RR}}
=
2\chi_{kk}^{\mathrm{RR}}
\xrightarrow[L\to\infty]{}
\frac{1}{2\sinh^2 g}.
\end{equation}
Additionally, we have
\begin{equation}
F_{kg}^{\mathrm{RR}}
=
\partial_k\!\left( \ii \langle u^{\rm R} \rvert \partial_g u^{\rm R} \rangle \right)
-
\partial_g\!\left( \ii \langle u^{\rm R} \rvert \partial_k u^{\rm R} \rangle \right)
=
\partial_g \langle n \rangle_g .
\end{equation}
Thus, $F_{kg}^{\mathrm{RR}}$ measures the response of the center of the right eigenstate to the non-Hermitian deformation $g$, and it is therefore naturally tied to the emergence of the skin localization.

\subsubsection{Biorthogonal quantum geometric tensor}

We next consider the quantum geometric tensor constructed from both right and left eigenstates:
\begin{equation}
\chi_{\mu\nu}^{\mathrm{LR}}
\coloneqq
\langle\!\langle \partial_\mu u^{\rm L} \rvert
\bigl(1-\lvert u^{\rm R}\rangle \langle\!\langle u^{\rm L} \rvert\bigr)
\lvert \partial_\nu u^{\rm R} \rangle,
\quad
\mu,\nu \in \{k,g\}.
\label{eq:gLR_HN}
\end{equation}
Since we have
\begin{equation}
\partial_g \lvert u^{\rm R} \rangle
=
-\ii\,\partial_k \lvert u^{\rm R} \rangle,
\quad
\partial_g \lvert u^{\rm L} \rangle\!\rangle
=
+
\ii\,\partial_k \lvert u^{\rm L} \rangle\!\rangle,
\end{equation}
we obtain
\begin{align}
\chi^{\mathrm{LR}}
&=
\chi_{kk}^{\mathrm{LR}}
\begin{pmatrix}
1 & -\ii \\
- \ii & - 1
\end{pmatrix}, \\
\chi_{kk}^{\mathrm{LR}}
&=
\frac{1}{L}\sum_{n=1}^{L} n^2
-
\left(
\frac{1}{L}\sum_{n=1}^{L} n
\right)^2
=
\frac{L^2-1}{12}.
\end{align}
In contrast to the right-right tensor, $\chi^{\mathrm{LR}}$ is completely independent of $g$. 
This is because $\ket{u^{\rm R}}$ and $| u^{\rm L} \rangle\!\rangle$ are localized at opposite boundaries.
Therefore, the biorthogonal quantum metric does not detect the skin localization scale.

The corresponding biorthogonal Berry curvature also vanishes:
\begin{equation}
F_{kg}^{\mathrm{LR}}
=
\partial_k\!\left( \ii \langle\!\langle u^{\rm L} \rvert \partial_g u^{\rm R} \rangle \right)
-
\partial_g\!\left( \ii \langle\!\langle u^{\rm L} \rvert \partial_k u^{\rm R} \rangle \right)
=
0,
\end{equation}
which is consistent with the absence of any $g$ dependence in $\chi^{\mathrm{LR}}$. We also note that, unlike in the right-right construction, the relation
\begin{equation}
F_{kg}^{\mathrm{LR}} = -2\,\operatorname{Im}\chi_{kg}^{\mathrm{LR}}
\end{equation}
does not generally hold.

In summary, the Hatano-Nelson model provides a simple and transparent setting in which two distinct notions of quantum geometry lead to qualitatively different conclusions. 
The quantum geometric tensor defined solely from right eigenstates faithfully captures the localization length scale induced by the skin effect, whereas the biorthogonal quantum geometric tensor does not. 
This contrast suggests that the physically relevant notion of quantum geometry in non-Hermitian systems depends crucially on how the underlying eigenstates are used to construct it.
In the presence of the skin effect, the plane-wave contributions should exhibit behavior similar to the Hatano-Nelson model even in generic non-Hermitian systems (see also Sec.~\ref{sec: plane-wave vs internal}).
Thus, we hereafter focus on the internal contributions to the quantum metric in multiband non-Hermitian systems, especially the non-Hermitian Su-Schrieffer-Heeger model.

\section{Non-Hermitian Su-Schrieffer-Heeger model}
    \label{sec: NH-SSH}

Next, we consider a non-Hermitian, nonreciprocal extension of the Su-Schrieffer-Heeger model~\cite{SSh-79}.
This paradigmatic two-band model has played a central role in the study of the non-Hermitian skin effect and the restoration of the bulk-boundary correspondence~\cite{Lee-16, YW-18-SSH, Kunst-18, Yokomizo-19}.
Here, we investigate the quantum metric
associated with its Bloch and non-Bloch eigenstates, and compare the metric constructed solely from right eigenstates with the biorthogonal metric defined from both right and left eigenstates. 
We show that both quantities diverge as the energy gap closes, and that their critical singularities are qualitatively different.
Notably, as a consequence of the non-Hermitian skin effect, the gap-closing points depend on the different boundary conditions.
We demonstrate that the quantum metrics can identify the gap-closing points appropriate to each boundary condition.
While we focus on the simple case in this section, we shortly study more generic cases where the generalized Brillouin zone exhibits singularities in Sec.~\ref{sec: non-Bloch}.

\subsection{Model}

We consider the non-Hermitian Su-Schrieffer-Heeger Hamiltonian~\cite{Lee-16, YW-18-SSH, Kunst-18, Yokomizo-19}, whose Bloch Hamiltonian is given as
\begin{equation}
H(k)=
\begin{pmatrix}
0 & p(k) \\
q(k) & 0
\end{pmatrix},
\end{equation}
with
\begin{align}
p(k) &= t_1+\frac{\gamma}{2}+t_2 e^{-\ii k}, 
\\
q(k) &= t_1-\frac{\gamma}{2}+t_2 e^{\ii k}.
\end{align}
While $t_1 \in \mathbb{R}$ and $t_2 \in \mathbb{R}$ describe the Hermitian hopping amplitudes, $\gamma \in \mathbb{R}$ describes the non-Hermitian, nonreciprocal hopping amplitude.
The corresponding eigenenergies are
\begin{align}
E_\pm(k) &=\pm \sqrt{p(k)\,q(k)} \nonumber \\
&=\pm \sqrt{\left( t_1 + t_2 \cos k \right)^2+\left( t_2 \sin k+\frac{\ii\gamma}{2}\right)^2}.
\label{eq:nhssh_energy}
\end{align}
In what follows, we focus on the lower band,
\begin{equation}
E_-(k)=-\sqrt{p(k)\,q(k)}.
\end{equation}
While Ref.~\cite{Ye-24} also investigated this model under the periodic boundary conditions, we here focus primarily on the open boundary conditions, highlighting the non-Hermitian skin effect.

It is convenient to introduce
\begin{equation}
z(k)\coloneqq\frac{p(k)}{q(k)}.
\label{eq:z_def}
\end{equation}
Then, a right eigenstate of the lower band is written as
\begin{equation}
\ket{u_-^{\mathrm R}(k)}
=
\frac{1}{\sqrt{1+|z(k)|}}
\begin{pmatrix}
-\sqrt{z(k)}\\
1
\end{pmatrix},
    \label{eq:right_eigenstate}
\end{equation}
where the square-root branch is chosen continuously away from singular points.
This state is normalized with respect to the usual inner product:
\begin{equation}
\braket{u_-^{\mathrm R}(k)|u_-^{\mathrm R}(k)}=1.
    \label{eq:right_normalization}
\end{equation}
The corresponding left eigenstate is given as
\begin{equation}
\langle\!\langle u_-^{\mathrm L}(k) |
=
\frac{\sqrt{1+|z(k)|}}{2}
\begin{pmatrix}
-\dfrac{1}{\sqrt{z(k)}} &
1
\end{pmatrix},
    \label{eq:left_eigenstate}
\end{equation}
which satisfies the biorthonormality condition
\begin{equation}
\langle\!\langle u_-^{\mathrm L}(k) \ket{u_-^{\mathrm R}(k)}=1.
\label{eq:biorthonormality}
\end{equation}

\subsection{Quantum geometry}

\subsubsection{Metric defined from right eigenstates}

We first consider the quantum metric defined solely from the normalized right eigenstate:
\begin{equation}
\chi^{\mathrm{RR}}(k)
=
\braket{\partial_k u_-^{\mathrm R}(k)|\partial_k u_-^{\mathrm R}(k)}
-\left|\braket{u_-^{\mathrm R}(k)|\partial_k u_-^{\mathrm R}(k)}\right|^2.
    \label{eq:gRR_def}
\end{equation}
A straightforward calculation from Eq.~\eqref{eq:right_eigenstate} yields
\begin{align}
\chi^{\mathrm{RR}}(k)
&=
\frac{\left|\partial_k \sqrt{z(k)}\right|^2}{\left(1+|z(k)|\right)^2}
\nonumber\\
&=
\frac{|z(k)|}{4\left(1+|z(k)|\right)^2}
\left|\partial_k \log z(k)\right|^2
\nonumber\\
&=
\frac{1}{4}
\frac{\left|\dfrac{p(k)}{q(k)}\right|}{\left(1+\left|\dfrac{p(k)}{q(k)}\right|\right)^2}
\left|
\frac{\partial_k p(k)}{p(k)}-\frac{\partial_k q(k)}{q(k)}
\right|^2.
\label{eq:gRR_formula}
\end{align}
This quantity is real and nonnegative, and it is thus a natural analog of the
Fubini-Study metric for wave functions in Hermitian systems.

\subsubsection{Biorthogonal metric}

We next consider the biorthogonal metric defined from both right and left eigenstates:
\begin{equation}
\chi^{\mathrm{LR}}(k)
= \langle\!\langle \partial_k u_-^{\mathrm L}(k) |
\left(1-\ket{u_-^{\mathrm R}(k)} \langle\!\langle u_-^{\mathrm L}(k) | \right)
\ket{\partial_k u_-^{\mathrm R}(k)}.
\label{eq:gLR_def}
\end{equation}
Using Eqs.~\eqref{eq:right_eigenstate} and \eqref{eq:left_eigenstate}, we find
\begin{equation}
\chi^{\mathrm{LR}}(k)
=
-\frac{1}{16}
\left(
\frac{\partial_k p(k)}{p(k)}-\frac{\partial_k q(k)}{q(k)}
\right)^2.
\label{eq:gLR_formula}
\end{equation}
Unlike $\chi^{\mathrm{RR}}$, the quantity $\chi^{\mathrm{LR}}$ is generally complex
and is not positive definite. 
Nevertheless, it is invariant under the
$\mathrm{GL}(1,\mathbb{C})$ gauge transformation
\begin{equation}
\ket{u_-^{\mathrm R}(k)}\to e^{f (k)} \ket{u_-^{\mathrm R}(k)},
\quad
\langle\!\langle u_-^{\mathrm L}(k) | \to e^{-f (k)}\langle\!\langle u_-^{\mathrm L}(k) |,
\end{equation}
with an arbitrary complex function $f (k) \in \mathbb{C}$,
preserving the biorthonormality condition
$\langle\!\langle u_-^{\mathrm L}(k) \ket{u_-^{\mathrm R}(k)}=1$.

\subsection{Periodic boundary conditions}

Under the periodic boundary conditions, momentum is real (i.e., $k\in\mathbb{R}$), and the quantum metrics are given directly by
Eqs.~\eqref{eq:gRR_formula} and \eqref{eq:gLR_formula}.
The metrics diverge for either $p(k^*)=0$ or $q(k^*)=0$ for some $k^*\in[0,2\pi)$.
Since the band energies are given by
$E_\pm(k)=\pm\sqrt{p(k)q(k)}$, this condition is precisely the condition for
the energy gap to close. 
Thus, both $\chi^{\mathrm{RR}}$ and
$\chi^{\mathrm{LR}}$ detect the band-touching point, although they do so with
different singular behaviors, as shown below.

\subsubsection{Critical behavior of $\chi^{\mathrm{RR}}$}

Suppose that there exists $k^*\in[0,2\pi)$ such that
\begin{equation}
p(k^*)=0,
\qquad
q(k^*)\neq 0.
\end{equation}
In the generic case, we may expand
\begin{equation}
p(k)\simeq p'(k^*)(k-k^*) \quad \left( p'(k^*) \neq 0 \right).
    \label{eq:p_expand}
\end{equation}
Then, Eq.~\eqref{eq:gRR_formula} gives
\begin{equation}
\chi^{\mathrm{RR}}(k)
\simeq
\frac{|p'(k^*)|}{4|q(k^*)|}
\frac{1}{|k-k^*|}.
\end{equation}
Therefore, the right-right quantum metric diverges with a first-order pole in
the momentum deviation.
The same conclusion holds for $q(k^*)=0$ and $p(k^*)\neq 0$, with the roles
of $p$ and $q$ interchanged.

\subsubsection{Critical behavior of $\chi^{\mathrm{LR}}$}

Under the same assumption $p(k^*)=0$ and $q(k^*)\neq 0$, Eq.~\eqref{eq:gLR_formula} yields
\begin{align}
\mathrm{Re}\,\chi^{\mathrm{LR}}(k)
&\simeq
-\frac{1}{16}\frac{1}{(k-k^*)^2}, \\
\mathrm{Im}\,\chi^{\mathrm{LR}}(k)
&\simeq
\frac{1}{8} \mathrm{Im} \left[ \frac{q' \left( k^* \right)}{q \left( k^* \right)} \right] \frac{1}{k-k^*}.
\end{align}
Notably, the power-law divergence of $\mathrm{Re}\,\chi^{\mathrm{LR}}$ is independent of the detailed forms of $p (k)$ and $q (k)$.
The biorthogonal metric exhibits a second-order pole, which is more
singular than the first-order divergence of $\chi^{\mathrm{RR}}$.
This distinction is one of the central observations of the present analysis:
although both notions of quantum geometry are sensitive to the gap closing,
the critical exponents of their singularities are different.

\subsection{Open boundary conditions}

We next turn to the open boundary conditions. 
In non-Hermitian systems, the
appropriate bulk description is formulated not necessarily on the ordinary Brillouin zone,
but on the generalized Brillouin zone, where momentum is complex valued.
According to the non-Bloch band theory~\cite{YW-18-SSH, Yokomizo-19}, 
we have in the present model
\begin{equation}
\left|e^{\ii k}\right|
= b \coloneqq
\sqrt{
\left|
\frac{t_1-\gamma/2}{t_1+\gamma/2}
\right|
}.
    \label{eq:gbz_radius}
\end{equation}
We therefore introduce
\begin{equation}
\beta\coloneqq e^{\ii k}= b e^{\ii\theta},
\qquad
\theta\in[0,2\pi),
    \label{eq:beta_theta}
\end{equation}
and regard $\theta$ as a natural momentum variable on the generalized Brillouin zone.
We thus consider the quantum geometry with respect to $\theta$.
While $b$ is a constant in the present case, it generally depends on $\theta$, as discussed in Sec.~\ref{sec: non-Bloch}.

In terms of $\beta$, the functions $p$ and $q$ become (for $t_3 = 0$)
\begin{align}
p(\beta)
&=
t_1+\frac{\gamma}{2}+t_2\beta^{-1}
=
t_1+\frac{\gamma}{2}+\frac{t_2}{b}e^{-\ii\theta},
\label{eq:p_beta}
\\
q(\beta)
&=
t_1-\frac{\gamma}{2}+t_2\beta
=
t_1-\frac{\gamma}{2}+t_2 b e^{\ii\theta}.
\label{eq:q_beta}
\end{align}
Then, we consider the quantum metrics under the open boundary conditions by replacing $k$ with $\theta$ and
$p(k),q(k)$ with $p(\beta),q(\beta)$:
\begin{align}
\chi^{\mathrm{RR}}(\theta)
&=
\frac{1}{4}
\frac{\left|\dfrac{p(\beta)}{q(\beta)}\right|}
{\left(1+\left|\dfrac{p(\beta)}{q(\beta)}\right|\right)^2}
\left|
\frac{\partial_\theta p(\beta)}{p(\beta)}
-
\frac{\partial_\theta q(\beta)}{q(\beta)}
\right|^2,
    \label{eq:gRR_obc}
\\
\chi^{\mathrm{LR}}(\theta)
&=
-\frac{1}{16}
\left(
\frac{\partial_\theta p(\beta)}{p(\beta)}
-
\frac{\partial_\theta q(\beta)}{q(\beta)}
\right)^2.
    \label{eq:gLR_obc}
\end{align}
Note that $p (\beta)$ and $q (\beta)$ are now functions of $\theta$ [see Eqs.~\eqref{eq:p_beta} and \eqref{eq:q_beta}],
and that $\partial_\theta p (\beta)$ and $\partial_\theta q (\beta)$ should be understood as
\begin{equation}
    \partial_\theta p (\beta) = \frac{dp}{d\beta} \frac{d\beta}{d\theta},\quad \partial_\theta q (\beta) = \frac{dq}{d\beta} \frac{d\beta}{d\theta}.
\end{equation}

These metrics diverge when we have either $p(\beta^*)=0$ or $q(\beta^*)=0$ for some
$\beta^*$ on the generalized Brillouin zone. In the present two-band model,
this is precisely the condition for the band gap under the open boundary conditions to close, because the non-Bloch band energies are again given by
\begin{equation}
E_\pm(\beta)=\pm\sqrt{p(\beta)\,q(\beta)}.
\end{equation}
Thus, the same statement holds as in the periodic boundary conditions: 
the quantum metrics generally diverge at the gap-closing points under the open boundary conditions, but their critical singularities are different for the choice of $\chi^{\mathrm{RR}}$ or $\chi^{\mathrm{LR}}$.

\subsection{Discussion}

The above analysis shows that the divergence of the quantum metric generally corresponds to the closure of the band gap, although the relevant gap-closing point depends on the boundary conditions. 
Under the periodic boundary conditions, the singularity appears when the ordinary Bloch Hamiltonian becomes gapless for a real momentum.
Under the open boundary conditions, by contrast, the singularity instead appears when the non-Bloch Hamiltonian becomes gapless for a complex momentum on the generalized Brillouin zone.
The gap closing does not, however, necessarily cause the divergence of the quantum metrics.
In particular, when $p$ and $q$ vanish simultaneously, the quantum metrics remain finite, as shown in Sec.~\ref{sec: cancellation}.
In passing, we note that Eqs.~\eqref{eq:p_beta} and \eqref{eq:q_beta} realize precisely this situation.

Given that gap closing can cause the divergence of the quantum metric even in Hermitian systems,
an important consequence of non-Hermiticity is that the right-right and left-right constructions encode different aspects of non-Hermitian criticality. 
The right-right metric
$\chi^{\mathrm{RR}}$ remains a real, nonnegative quantity and generally exhibits a first-order pole.
By contrast, the biorthogonal quantity $\chi^{\mathrm{LR}}$ is generally complex and
develops a second-order pole. 
In this sense, while both quantities detect the same gap-closing conditions, they do not characterize them in the same manner.
We also note that the gapless points in the non-Hermitian Su-Schrieffer-Heeger model are exceptional points.
Nevertheless, the singularities of the quantum metric at them are similar to those at gapless points in Hermitian systems.

\section{Non-Bloch band theory}
    \label{sec: non-Bloch}

We now consider the extent to which the results obtained in Sec.~\ref{sec: NH-SSH} persist in more general non-Hermitian Su-Schrieffer-Heeger-type models. 
For example, upon including additional
hopping terms, the trajectory of complex momentum, namely, the generalized Brillouin zone, can acquire a more
intricate structure, possibly with nonanalytic points. 
In this section, we demonstrate that such singular features are also encoded in the quantum geometry.
Specifically, we show that cusps on the generalized Brillouin zone (see Fig.~\ref{GBZ-cusp}) are manifested as discontinuities in the quantum geometry.

\subsection{Generalized Brillouin zone}

\begin{figure}[t]
\includegraphics[width=1.0\linewidth]{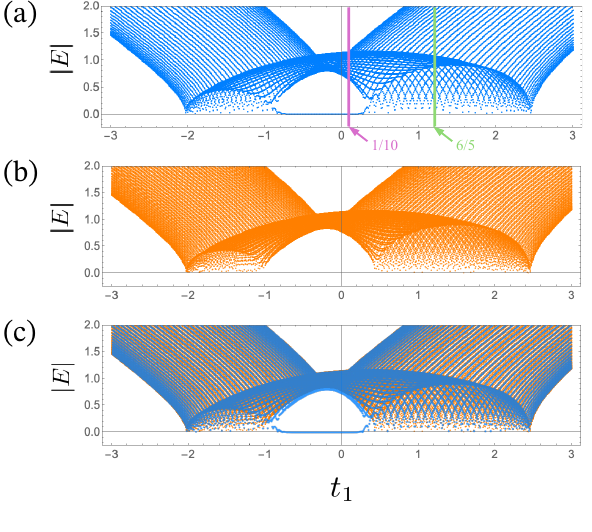}
\caption{Complex spectrum of the generalized non-Hermitian Su-Schrieffer-Heeger model ($t_2 = 1$, $t_3 = 2/5$, $\gamma_1 = 2$, $\gamma_2 = 2/3$) [Eq.~\eqref{param1}] under the open boundary conditions as a function of $t_1$.
(a)~Complex spectrum obtained from the direct diagonalization for the system size $L=100$.
(b)~Complex spectrum obtained from the generalized Brillouin zone.
The last panel~(c) shows both~(a) (blue) and (b) (orange).}
\label{spec-obc}
\end{figure}

To characterize the quantum geometry on the generalized Brillouin zone, we investigate the generalized non-Hermitian Su-Schrieffer-Heeger model~\cite{Lee-16, YW-18-SSH, Kunst-18, Yokomizo-19}:
\begin{equation}
H(\beta)=
\begin{pmatrix}
0 & p(\beta) \\
q(\beta) & 0
\end{pmatrix},
    \label{SSH-bloch}
\end{equation}
with
\begin{align}
p(\beta) &= t_1+\frac{\gamma_1}{2}+ \left( t_2 - \frac{\gamma_2}{2} \right) \beta^{-1} + t_3 \beta, \label{p} \\
q(\beta) &= t_1-\frac{\gamma_1}{2} + \left( t_2 + \frac{\gamma_2}{2} \right) \beta + t_3 \beta^{-1}. \label{q}
\end{align}
Here, $t_1, t_2, t_3 \in \mathbb{R}$ represent the Hermitian terms, whereas $\gamma_1, \gamma_2 \in \mathbb{R}$ denote the non-Hermitian terms.
This model reduces to the model in Sec.~\ref{sec: NH-SSH} upon setting $\gamma_1 = \gamma$ and $\gamma_2 = t_3 = 0$.
Under the periodic boundary conditions, $\beta$ lies on the unit circle, i.e., $\beta = e^{\ii k}$ with $k \in \left[ 0, 2\pi \right)$.
However, under the open boundary conditions, $\beta$ generally departs from the unit circle and traces a generic loop in the complex plane, thereby forming the generalized Brillouin zone~\cite{YW-18-SSH, Yokomizo-19}.
The deviation from the unit circle, particularly $\mathrm{Re} \log \beta$, quantifies the localization of skin modes.
While the model in Sec.~\ref{sec: NH-SSH} yields only a simple circle of the generalized Brillouin zone [i.e., Eqs.~\eqref{eq:gbz_radius} and \eqref{eq:beta_theta}], the model in Eq.~\eqref{SSH-bloch} can generally form a more intricate loop structure, as discussed in the following.

As the parameters $t_{1,2,3}$ and $\gamma_{1,2}$ are varied,
the model exhibits different types of complex spectra under the open boundary conditions.
Figure~\ref{spec-obc}\,(a) gives examples of the complex spectrum under the open boundary conditions for 
\begin{equation}
t_2 = 1, \quad t_3 = \frac{2}{5}, \quad \gamma_1 = 2, \quad \gamma_2 = \frac{2}{3}
    \label{param1}
\end{equation}
as varying $t_1$, obtained by the direct diagonalization for the system size $L=100$.
For example, for $t_1=1/10$ 
(indicated in the figure; this choice of $t_1$ will be employed in the subsequent calculations of the quantum metric) and in a parameter range of $t_1$ around it,
the system lies in the topologically nontrivial phase hosting a pair of zero-energy edge states.
By contrast, in Fig.~\ref{spec-obc}\,(a),
$t_1=6/5$ (also indicated in the figure)
belongs to another typical regime of parameters of the model where the spectrum becomes gapless,
with a rather pronounced finite-size effect.

To incorporate the non-Hermitian skin effect into the quantum metric,
we here employ the generalized Brillouin zone approach~\cite{YW-18-SSH, Yokomizo-19}.
A significant advantage of this approach is that it allows us to reproduce the bulk properties of the model under the open boundary conditions
from the Bloch-like Hamiltonian in Eq.~\eqref{SSH-bloch},
provided that the conventional Brillouin zone is replaced by the generalized Brillouin zone.
In this formulation, we have $\beta=e^{\ii k}$,
where $k \in \mathbb{C}$ is a generalized complex-valued wave number,
or equivalently, $\beta$ acquires a nontrivial amplitude, $b \coloneqq |\beta| \neq 1$.
In the prescription of the generalized Brillouin zone~\cite{Yokomizo-19},
$\beta$ is determined as a collection of points
that forms a trajectory in the complex-$\beta$ plane, namely, the generalized Brillouin zone:
\begin{equation}
\beta = \beta(\theta) = b(\theta) e^{\ii\theta} \quad \left( b > 0, 0 \leq \theta < 2\pi \right).
    \label{GBZ-def}
\end{equation}
Notably, the amplitude  $b(\theta)=|\beta(\theta)|$ not only deviates from unity but also depends explicitly on $\theta = \arg\beta$.

\begin{figure}[t]
\includegraphics[width=0.6\linewidth]{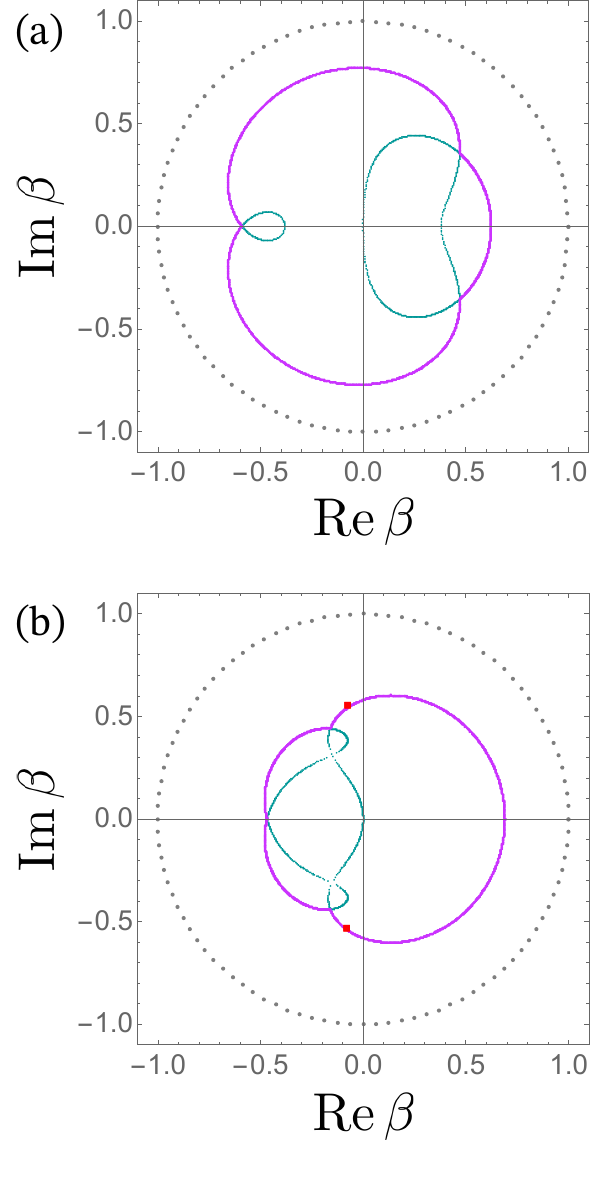}
\caption{Generalized Brillouin zone for (a)~$t_1=1/10$ 
and (b)~$t_1=6/5$ 
(purple plots).
The other parameters of the model are chosen as in Eq.~\eqref{param1}.
For (b), the energy spectrum has a pair of gapless points,
which correspond to the red squares on the generalized Brillouin zone.}
    \label{GBZ-cusp}
\end{figure}

In Fig.~\ref{GBZ-cusp}, we show the generalized Brillouin zone for some specific cases, i.e., the trajectory of $\beta = \beta(\theta)$ in the complex $\beta$-plane,
obtained in the form of Eq.~\eqref{GBZ-def}.
The parameters are chosen as $t_1 = 1/10$ for Fig.~\ref{GBZ-cusp}\,(a) and $t_1 = 6/5$ for Fig.~\ref{GBZ-cusp}\,(b), 
both of which are also indicated in Fig.~\ref{spec-obc}\,(a),
while the other parameters are fixed as in Eq.~\eqref{param1}.
In Fig.~\ref{GBZ-cusp}, we also plot these trajectories of $\beta$ against the unit circle (represented by the gray dotted circle),
to emphasize that the magnitude $b(\theta)=|\beta(\theta)|$ deviates
generally from unity.
In contrast to the simple case discussed in Sec.~\ref{sec: NH-SSH}, the generalized Brillouin zone hosts cusp singularities.
Specifically, each case shown in Fig.~\ref{GBZ-cusp} contains three such cusps.

\begin{figure}
\includegraphics[width=1.0\linewidth]{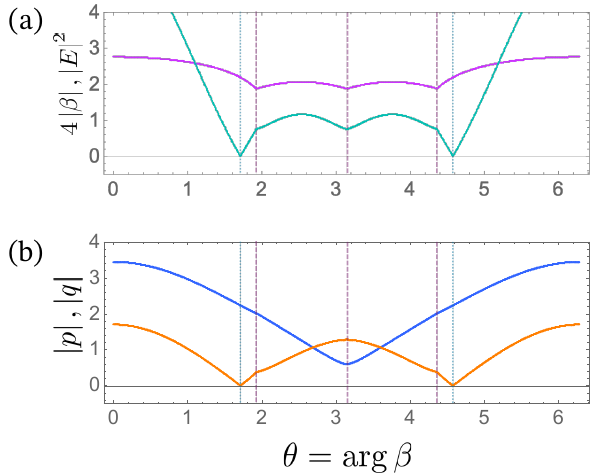}
\caption{Cusps and gapless points in the generalized Brillouin zone
for $t_1=6/5$ and Eq.~\eqref{param1}.
(a)~$4|\beta|$ 
(purple; rescaled by a factor of $4$ for visibility) and $|E|^2$ (light greenish blue) 
as a function of $\theta=\arg\beta$ on the generalized Brillouin zone.
(b)~$|p|$ (blue) and $|q|$ (orange).}
    \label{cusp-gapless}
\end{figure}

From the energy dispersions shown in Fig.~\ref{spec-obc}, one can already see that multiple branches of the dispersion relation overlap 
and collectively form the actual energy dispersion. 
For example, even when focusing only on the case with $t_1=1/10$, 
the spectrum is assembled from several overlapping branches. 
One can also observe that, at certain energies, 
the branch contributing to the energy dispersion switches from one branch to another. 
The value of $\beta$ corresponding to the energy for this branch switching corresponds to a cusp on the generalized Brillouin zone. 
Naturally, corresponding to the cusp on the generalized Brillouin zone, 
the energy dispersion also exhibits a cusp on the generalized Brillouin zone 
[cf.~Fig.~\ref{cusp-gapless}\,(a)].
These cusps are not, however, directly related to a topological phase transition.
At most, they merely play a role in forming the band structure that makes such a topological phase transition possible. 
Therefore, one cannot, for instance, infer the existence of a cusp merely from the occurrence of a topological phase transition. 
How, then, can this cusp in turn be detected? 
In this work, we propose that the quantum metric provides a useful diagnostic for identifying the cusp.

At these cusps, the generalized Brillouin zone switches from one branch of the solutions to the eigenvalue equation, $E^2=pq$, to another branch~\cite{Yokomizo-19}.
In Fig.~\ref{GBZ-cusp}, these different branches are shown as the dark bluish curves in the background of the generalized Brillouin zone to highlight the switching at each cusp.
Although they do not necessarily correspond to the generalized Brillouin zone, they are useful as the auxiliary Brillouin zone~\cite{Yang-20}.
Substituting the values of $\beta$ on the generalized Brillouin zone into the expressions for $p$ in Eq.~\eqref{p} and $q$ in Eq.~\eqref{q},
one obtains the spectrum from the eigenvalue equation $E^2 = pq$, 
as shown in Fig.~\ref{spec-obc}\,(b).
This construction reproduces the bulk or continuum part of the complex spectrum under the open boundary conditions,
except for the zero-energy states.

In passing, we note that nonanalytic points on the generalized Brillouin zone are not cusps in the strict mathematical sense, although they are often referred to as cusps in the literature on non-Bloch band theory (see, for example, Ref.~\cite{Yokomizo-19}).
More precisely, in the mathematics of algebraic curves, a cusp is defined as a singular point at which a moving point on the curve must reverse direction.
By contrast, a typical nonanalytic point on the generalized Brillouin zone consists of two different smooth local branches with different tangential directions, which is more properly classified as a node.
In other words, a cusp is characterized by a single singular local branch, whereas a node is characterized by two transverse local branches.
In this work, we find that a node on the generalized Brillouin zone gives rise to a discontinuity in the quantum metric.
Still, following the conventional terminology in the non-Bloch band theory literature, we here refer to such nonanalytic points on the generalized Brillouin zone as cusps.

\begin{figure}
\includegraphics[width=0.6\linewidth]{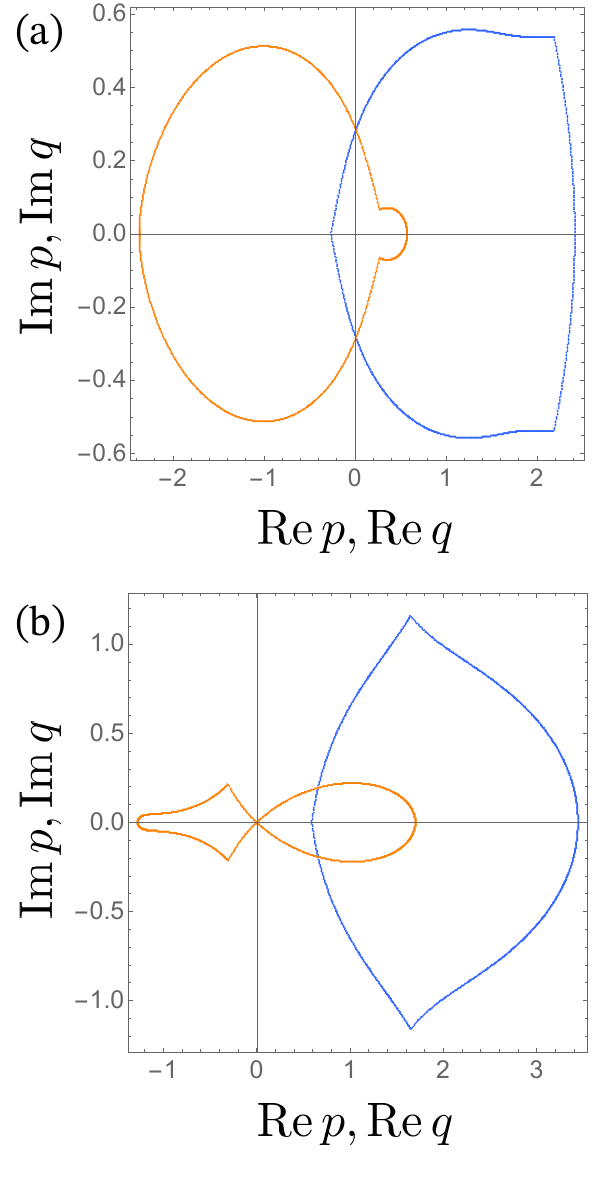}
\caption{Trajectory of $p=p(\beta)$ (blue) and $q=q(\beta)$ (orange) 
in the complex $p$- and $q$-planes
along the generalized Brillouin zone for (a)~$t_1=1/10$ and (b)~$t_1=6/5$.
The other parameters of the model are chosen as in Eq.~\eqref{param1}.
The nontrivial winding of $p(\beta)$ and $q(\beta)$ around the origin corresponds to the existence of a pair of the zero-energy edge modes~\cite{YW-18-SSH}.}
    \label{pq-GBZ}
\end{figure}
\begin{figure}
\includegraphics[width=1.0\linewidth]{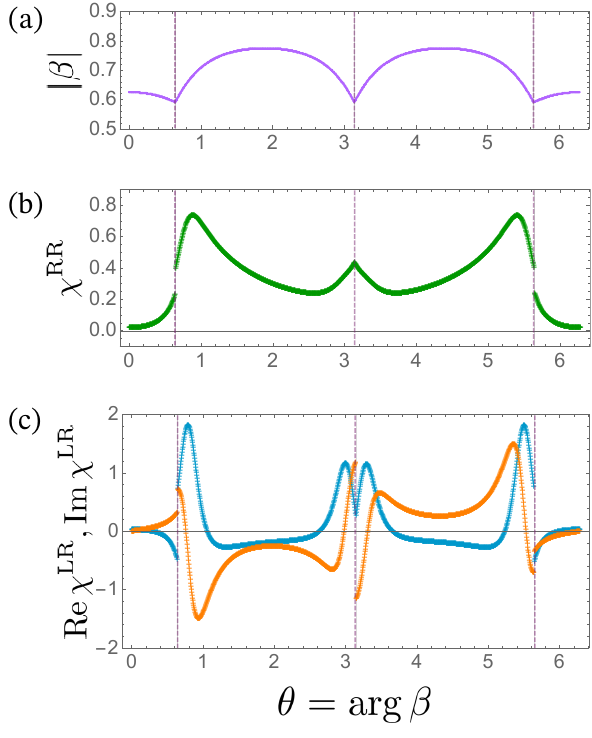}
\caption{Singular behaviors of the quantum metrics $\chi^{\rm RR}(\theta)$ and $\chi^{\rm LR} (\theta)$ at cusps of the generalized Brillouin zone.
The parameters are chosen for $t_1=1/10$ and Eq.~\eqref{param1}.
(a)~$\left| \beta \right|$ as a function of $\theta = \arg \beta$.
Cusps appear for $\theta \approx 0.64, \pi, 5.64$ 
(purple dashed lines).
(b)~$\chi^{\rm RR}(\theta)$.
(c)~$\mathrm{Re}\,\chi^{\rm LR}(\theta)$ (blue) and $\mathrm{Im}\,\chi^{\rm LR}(\theta)$ (orange).
}
\label{chi-cusp}
\end{figure}

\subsection{Cusps but no gapless points}

The trajectories of $p = p (\beta)$ and $q = q (\beta)$ as functions of $\beta$ on the generalized Brillouin zone determine the topological invariant of the non-Hermitian Su-Schrieffer-Heeger model~\cite{YW-18-SSH}.
In Fig.~\ref{pq-GBZ}\,(a),
we show these trajectories of $p(\beta)$ and $q(\beta)$ for $t_1=1/10$, with the remaining parameters chosen as in Eq.~\eqref{param1}.
These trajectories show that
both $p(\beta)$ and $q(\beta)$ wind around the origin 
in the complex $p$- and $q$-planes
as $\beta$ encircles the origin in the complex $\beta$-plane, namely, along the generalized Brillouin zone.
The difference between the winding numbers of $p(\beta)$ and $q(\beta)$ 
is in one-to-one correspondence 
with the existence of a pair of zero-energy edge modes, thereby providing the generalized bulk-boundary correspondence.
Plugging the same values of $\beta$ into Eqs.~\eqref{eq:gRR_obc} and \eqref{eq:gLR_obc}, we obtain the quantum metrics.
Notably, Eqs.~\eqref{eq:gRR_obc} and \eqref{eq:gLR_obc} involve derivatives of $p(\beta)$ and $q(\beta)$ with respect to $\theta$, 
which mean $\partial_\theta p \left( \beta \left( \theta \right) \right)$ and $\partial_\theta q \left( \beta \left( \theta \right) \right)$ [recall that $\beta$ itself depends on $\theta$; see Eq.~\eqref{GBZ-def}].
In this manner, we calculate the quantum metrics $\chi^{\rm RR} \left( \theta \right)$ and $\chi^{\rm LR} \left( \theta \right)$ along the generalized Brillouin zone.

At cusps of the generalized Brillouin zone [see, for example, Fig.~\ref{GBZ-cusp}\,(a)], 
these derivatives become discontinuous; 
the derivative taken from one side of a cusp
does not, in general, match that taken from the other side.
Following the above procedure, we calculate $\chi^{\rm RR} (\theta)$ and $\chi^{\rm LR} (\theta)$, and confirm such discontinuities, as shown in Fig.~\ref{chi-cusp}.
Here, $\theta$-derivatives are taken numerically, using the data points for $\beta$'s on the generalized Brillouin zone.
Therefore, the branch switching of the generalized Brillouin zone is not merely a mathematical feature of the complex-momentum parametrization, but also induces an abrupt change in the physically relevant band geometry of non-Bloch eigenstates.

To identify the cusps in the generalized Brillouin zone, we first replot
$b(\theta)=|\beta(\theta)|$ in Fig.~\ref{chi-cusp}\,(a) as a function of
$\theta=\arg\beta$.
The three cusps $\theta^{\times}_{0, 1, 2}$ in the generalized Brillouin zone of Fig.~\ref{GBZ-cusp} are
found at 
\begin{equation}
\theta^{\times}_0 = \pi, \quad 
\theta^{\times}_1 \approx 0.64, \quad
\theta^{\times}_2 \approx 5.64 = 2\pi - \theta^{\times}_1.
\end{equation}
In Fig.~\ref{chi-cusp}\,(b), we present $\chi^{\rm RR}$ as a function of $\theta$, 
exhibiting either a discontinuity at $\theta=\theta^{\times}_{1, 2}$
or a cusp at $\theta=\pi$.
In Fig.~\ref{chi-cusp}\,(c), we plot the real and imaginary parts of $\chi^{\rm LR} = \chi^{\rm LR} (\theta)$.
Note that, while $\chi^{\rm RR}$ is always real and nonnegative, $\chi^{\rm LR}$ is generally complex.
The real part (plots in blue) 
exhibits discontinuities at $\theta=\theta^{\times}_{1, 2}$ and a cusp at $\theta=\pi$,
while
the imaginary part (plots in orange) 
shows discontinuities 
even at $\theta=\pi$.
Figures~\ref{chi-cusp}\,(b) and (c)
thus demonstrate that
the metrics
$\chi^{\rm RR} (\theta)$ and $\chi^{\rm LR} (\theta)$ 
become singular, typically through discontinuities,
at the cusps of the generalized Brillouin zone.
The absence of such discontinuities for $\chi^{\rm RR}$ and $\mathrm{Re}\,\chi^{\rm LR}$ at $\theta = \pi$, in contrast to $\theta = \theta^{\times}_{1,2}$, 
originates from time-reversal symmetry.
Indeed, the generalized non-Hermitian Su-Schrieffer-Heeger model in Eq.~\eqref{SSH-bloch} respects time-reversal symmetry:
\begin{equation}
    \left( H \left( \beta \right) \right)^* = H \left( \beta^* \right).
\end{equation}
Then, as demonstrated in Sec.~\ref{sec: symmetry},
both $\chi^{\rm RR}$ and $\mathrm{Re}\,\chi^{\rm LR}$ are symmetric under $\theta \to -\theta$, which in turn prohibits the discontinuity at the symmetric point $\theta = \pi$ (and also $\theta = 0$).
By contrast, 
$\mathrm{Im}\,\chi^{\rm LR}$
is antisymmetric under $\theta \to -\theta$ and hence compatible with the discontinuity at $\theta = \pi$.

\begin{figure}
\includegraphics[width=0.9\linewidth]{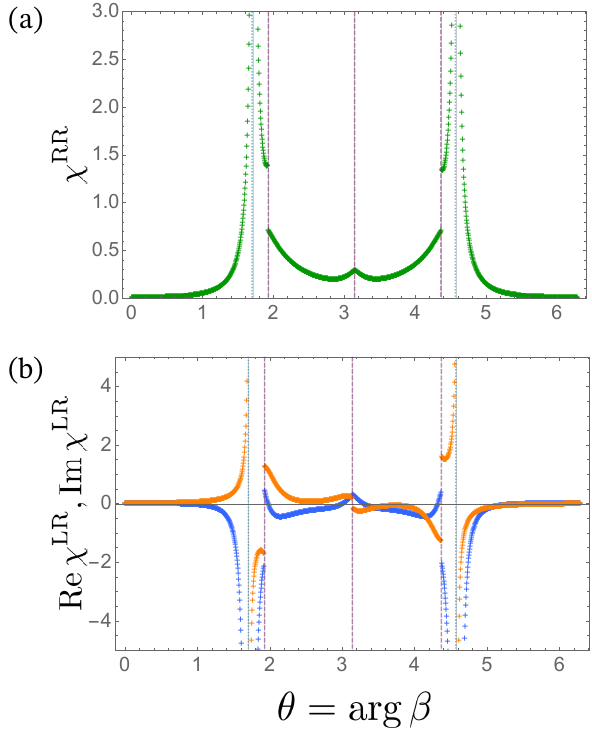}
\caption{Singular behaviors of the quantum metrics $\chi^{\rm RR}(\theta)$ and $\chi^{\rm LR} (\theta)$ along the generalized Brillouin zone.
The parameters are chosen for $t_1=6/5$ and Eq.~\eqref{param1}.
(a)~$\chi^{\rm RR}(\theta)$, exhibiting discontinuities at $\theta \approx 1.92, \pi, 4.36$ (purple dashed lines) 
and divergences at $\theta \approx 1.71, 4.57$ (light blue dotted lines).
(b)~$\mathrm{Re}\,\chi^{\rm LR}(\theta)$  (blue) 
and $\mathrm{Im}\,\chi^{\rm LR}(\theta)$ (orange).
}
\label{chi-gapless}
\end{figure}

\begin{figure*}
\includegraphics[width=0.75\linewidth]{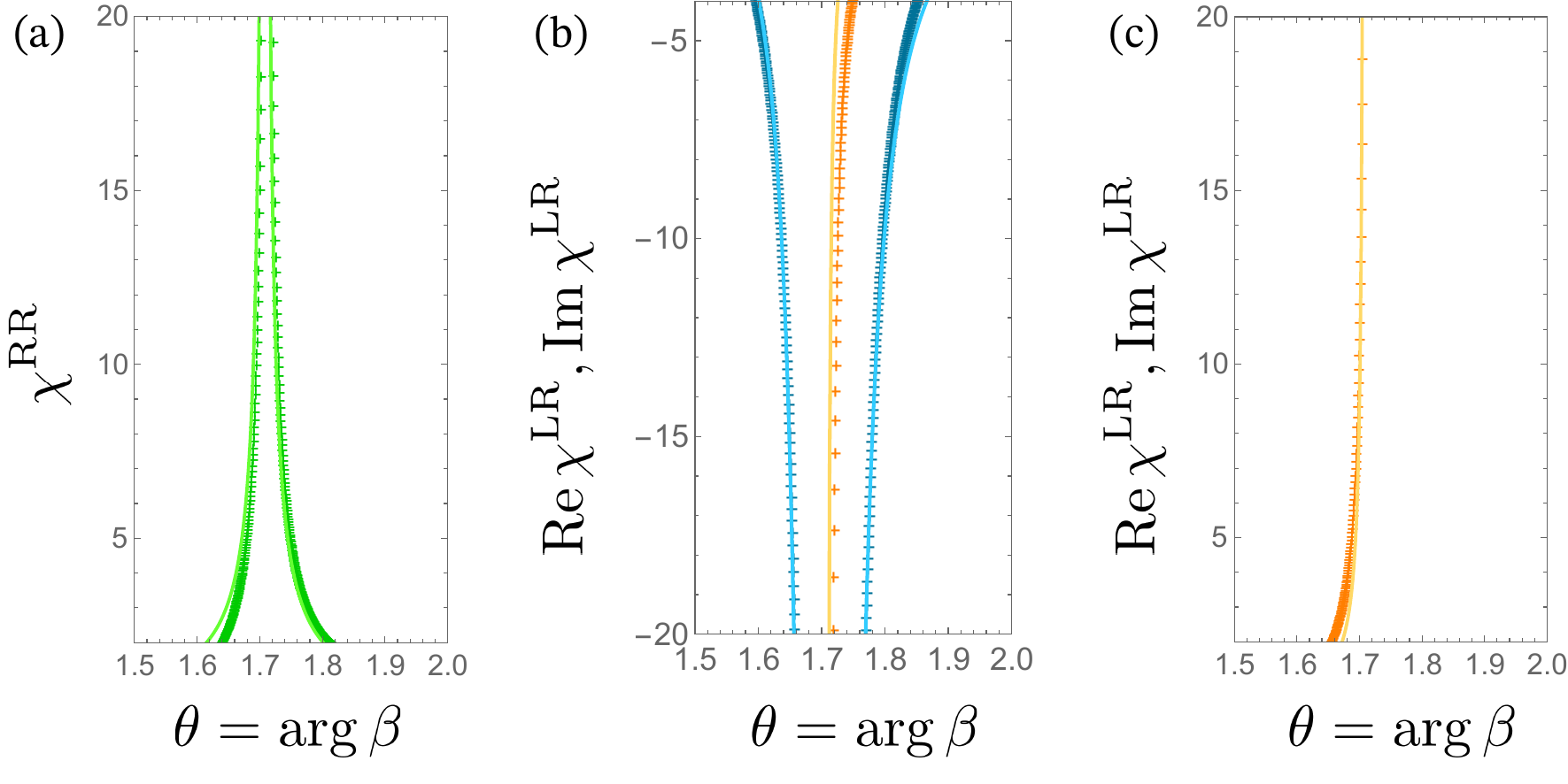}
\caption{Divergent behavior of (a)~$\chi^{\rm RR}$ (green), (b, c)~$\mathrm{Re}\,\chi^{\rm LR}$ (blue), and $\mathrm{Im}\,\chi^{\rm LR}$ (orange) in the vicinity of the gapless point $\theta = \theta_1^* \approx 1.71$.
These plots are similar to those in Fig.~\ref{chi-gapless}, but are obtained with improved numerical precision using
$12538$ $\beta$-points on the generalized Brillouin zone.
On top of the numerical data,
the asymptotic expressions in Eqs.~\eqref{chi_RR_asymp} [green curve in (a)], \eqref{chi_LR_asymp1} [blue curve in (b, c)], and  \eqref{chi_LR_asymp2} [orange curve in (b, c)] are superposed.}
    \label{chi-asympt}
\end{figure*}

\begin{figure}
\includegraphics[width=0.7\linewidth]{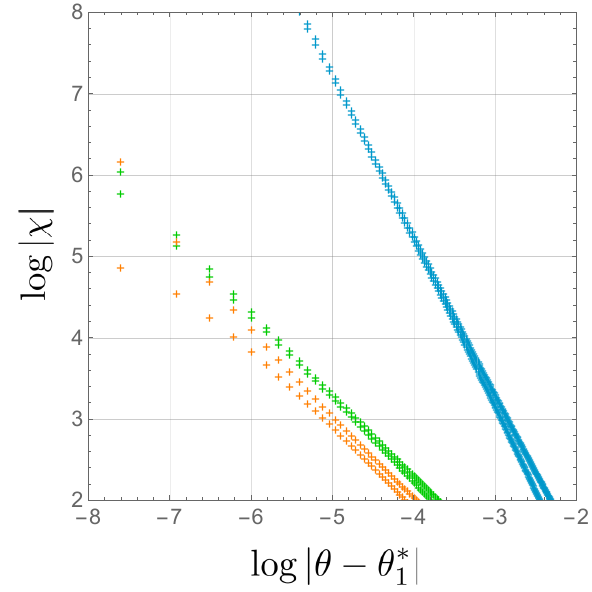}
\caption{Log-log plot of the divergent behavior of the quantum metrics $\left| \chi \right|$'s
as functions of $|\theta-\theta_1^*|$,
for $\chi^{\rm RR}$ (green),  
$\mathrm{Re}\,\chi^{\rm LR}$ (blue),
and $\mathrm{Im}\,\chi^{\rm LR}$ (orange).
The same $\beta$-points are used as in Fig.~\ref{chi-asympt}.
The slopes of the asymptotes, determining the power-law divergence [see Eq.~\eqref{alpha}] in the vicinity of the gapless point
$\theta=\theta_1^{*}$,
are obtained as $-\alpha \approx -1$ for $\chi^{\rm RR}$ (green) and $\mathrm{Im}\,\chi^{\rm LR}$ (orange), and $-\alpha \approx -2$ for $\mathrm{Re}\,\chi^{\rm LR}$ (blue).}
    \label{chi-log}
\end{figure}

\subsection{Cusps and a pair of gapless points}
    \label{sec: gapless}

Let us now turn to the case of $t_1=6/5$,
indicated in Fig.~\ref{spec-obc}\,(a) and 
corresponding to
Fig.~\ref{GBZ-cusp}\,(b) and Fig.~\ref{pq-GBZ}\,(b).
For this choice of the parameters,
while the trajectory of $\beta(\theta)$ shows cusps, the corresponding spectrum becomes gapless.
Indeed, the spectrum possesses a pair of gapless points on the generalized Brillouin zone [Fig.~\ref{cusp-gapless}\,(a)], which corresponds to a pair of zeros of $q(\beta)$.
Notably, the locations of the cusps do not generally coincide with those of the gapless points.
The cusps appear at $\theta = \theta^\times_{0, 1, 2}$
with 
\begin{equation}
\theta^\times_{0} = \pi, \quad
\theta^\times_{1} \approx 1.92, \quad
\theta^\times_{2} \approx 4.36 = 2\pi-\theta^\times_{1};
\end{equation}
see Fig.~\ref{GBZ-cusp}\,(b) and Fig.~\ref{cusp-gapless}\,(a).
On the other hand, the gapless points of the spectrum are located at
\begin{equation}
\theta^*_{1} \approx 1.71, \quad
\theta^*_{2} \approx 4.57 = 2\pi - \theta^*_{1}.
    \label{zeros}
\end{equation}
These two gapless points coincide with a pair of zeros of $q(\beta)$.
In Fig.~\ref{pq-GBZ}\,(b), we show the
trajectories of $p(\beta)$ (plotted in blue) and $q(\beta)$ (plotted in orange)
on the generalized Brillouin zone.
They indicate that the trajectory of $q(\beta)$ crosses itself at the origin,
implying the presence of a pair of zeros of $q(\beta)$.
In Fig.~\ref{chi-gapless}, we present the singular behaviors of $\chi^{\rm RR}(\theta)$ and $\chi^{\rm LR} (\theta)$
at these cusps and gapless points.
As noted before, $\chi^{\rm RR}(\theta)$ is always real, 
and here shows discontinuities at $\theta = \theta_{1,2}^\times$ and a cusp 
at $\theta=\pi$,
while it diverges at the gapless points $\theta = \theta_{1,2}^*$.

Recall that $q(\beta)$ vanishes at these gapless points,
i.e., $q(\beta_1)=q(\beta_2)=0$, 
where $\beta_{1,2} \coloneqq \beta(\theta_{1,2}^*)$.
On the contrary, $p(\theta) = p|_{\beta=\beta(\theta)}$
remains nonzero even at
these gapless points $\theta=\theta_{1,2}^*$. 
At a gapless point where either $p$ or $q$ vanishes, 
the quantum metrics $\chi^{\rm RR}(\theta)$ and $\chi^{\rm LR}(\theta)$ should diverge, while the critical behavior depends on the choice of the metric, as also discussed in Sec.~\ref{sec: NH-SSH}.
Assuming that $q(\theta)=q|_{\beta=\beta(\theta)}$ can be expanded around one of the zeros
$q(\theta_n^*)=0$ ($n=1,2$) as
\begin{equation}
q(\theta)\simeq  
q'(\theta_n^*)(\theta-\theta_n^*) + O (\theta-\theta_n^*)^2,
\end{equation}
where 
\begin{align}
q'(\theta)&= \frac{d}{d\theta} \left[q|_{\beta=\beta(\theta)}\right]
\nonumber \\
&=
\left( \frac{b'(\theta)}{b(\theta)} + \ii
\right)
\left. \frac{dq}{d\beta} \right|_{\beta=\beta(\theta)}
\beta(\theta),
\end{align}
we obtain
from Eq.~\eqref{eq:gRR_obc}
the asymptotic form of $\chi^{\rm RR}(\theta)$ as
\begin{equation}
\chi^{\rm RR}(\theta)\simeq
\frac{1}{4} \left|
\frac{q'(\theta^*_n)}{p(\theta^*_n)}
\frac{1}{\theta-\theta^*_n}
\right|.
    \label{chi_RR_asymp}
\end{equation}
Thus,
$\chi^{\rm RR}(\theta)$ diverges as $1/|\theta-\theta^*_n|$
around $\theta=\theta^*_n$.
Similarly, from Eq.~\eqref{eq:gLR_obc}, we obtain the asymptotic behavior of $\chi^{\rm LR} (\theta)$.
The real part of $\chi^{\rm LR} (\theta)$
around $\theta=\theta^*_n$ obeys
\begin{equation}
\mathrm{Re}\,\chi^{\rm LR}(\theta)\simeq
 - \frac{1}{16} \frac{1}{(\theta-\theta^*_n)^2},
    \label{chi_LR_asymp1}
\end{equation}
and hence diverges as $-1/(\theta-\theta^*_n)^2$.
By contrast, the imaginary part of $\chi^{\rm LR} (\theta)$ follows
\begin{equation}
\mathrm{Im}\,\chi^{\rm LR} (\theta) \simeq
\frac{1}{8} {\rm Im}
\left[ \frac{p'(\theta^*_n)}{p(\theta^*_n)}
\right]
\frac{1}{\theta-\theta^*_n},
\label{chi_LR_asymp2}
\end{equation}
diverging as $1/(\theta-\theta^*_n)$, unlike the real part.

In Fig.~\ref{chi-asympt},
we replot the divergent behavior of $\chi^{\rm RR}$ and $\chi^{\rm LR}$
shown in Fig.~\ref{chi-gapless},
with improved numerical precision using $12538$ $\beta$-points.
To evaluate $\chi^{\rm RR}$ and $\chi^{\rm LR}$,
we compute numerical derivatives on the basis of these $\beta$-points.
Then, on top of the numerical data,
we superpose the asymptotic expressions in Eqs.~\eqref{chi_RR_asymp}, \eqref{chi_LR_asymp1}, and \eqref{chi_LR_asymp2}.
These asymptotic curves are shown as the solid curves,
whereas the numerical data are plotted with the markers ``$+$",
both of which are in agreement.
These results confirm that the critical behavior at gapless points of the generalized Brillouin zone depends qualitatively on the type of the quantum metrics under consideration.
To further verify these power laws, we also present a log-log plot of
$\chi(\theta)$'s around a gapless point $\theta=\theta_1^*$, as shown in Fig.~\ref{chi-log}.
In this plot, the slope of the asymptotes determines the exponent $\alpha$
that controls the divergence of $\chi(\theta)$ as
\begin{equation}
\chi(\theta) \propto \frac{1}{|\theta-\theta_1^{*}|^\alpha}.
\label{alpha}
\end{equation}
From the plots in Fig.~\ref{chi-log}, we can deduce
$\alpha\simeq 1$ for $\chi^{\rm RR}$ and
$\mathrm{Im}\,\chi^{\rm LR}$, while $\alpha\simeq 2$ for $\mathrm{Re}\,\chi^{\rm LR}$, 
all of which are fully consistent with the asymptotic expressions in
Eqs.~\eqref{chi_RR_asymp}, \eqref{chi_LR_asymp1}, and \eqref{chi_LR_asymp2}.

\begin{figure}
\includegraphics[width=1.0\linewidth]{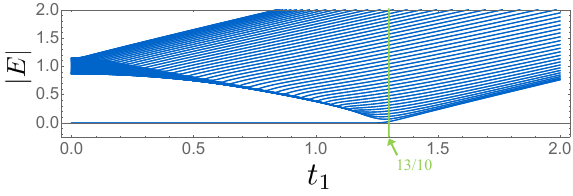}
\caption{Complex energy spectrum under the open boundary conditions as a function of $t_1$ for the model parameters $t_2 = 1$, $t_3 = 1/5$, $\gamma_1 = 2/3$, $\gamma_2 = 0$ [Eq.~\eqref{param2}].
$t_1=13/10$ (light green line) corresponds to a phase transition between topologically nontrivial and trivial phases.}
\label{spec-obc-gapped}
\end{figure}

\begin{figure}
\includegraphics[width=0.85\linewidth]{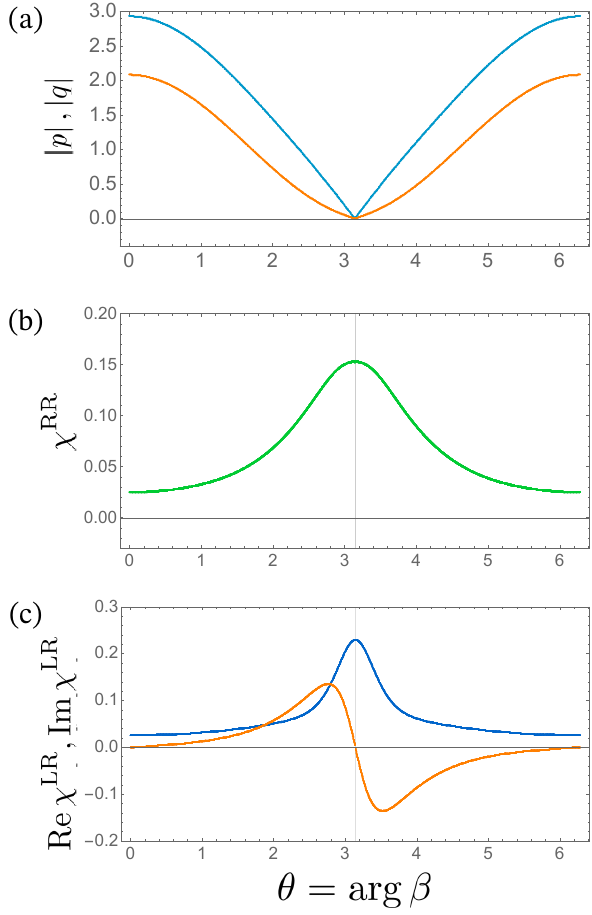}
\caption{Cancellation of divergences in (b)~$\chi^{\rm RR}(\theta)$ and (c)~$\chi^{\rm LR}(\theta)$ for $t_1=13/10$ and Eq.~\eqref{param2}.
(a)~$\left| p(\theta) \right|$ (blue) and $\left| q(\theta) \right|$ (orange) along the generalized Brillouin zone, 
both of which vanish at $\theta=\pi$.
(b)~$\chi^{\rm RR}$ as a function of $\theta = \arg \beta$.
(c)~$\mathrm{Re}\,\chi^{\rm LR}$ (blue) and $\mathrm{Im}\,\chi^{\rm LR}$ (orange).}
    \label{chi-cancel}
\end{figure}

\subsection{Cancellation of the divergence}
    \label{sec: cancellation}

While we have so far assumed that either $p$ or $q$ vanishes at a gapless point,
there are cases in which both $p$ and $q$ vanish simultaneously.
In view of the expressions of the quantum metrics in Eqs.~\eqref{eq:gRR_obc} and \eqref{eq:gLR_obc}, 
one may expect the corresponding critical behavior to be modified.
In this section, we demonstrate that this is indeed the case, and that the quantum metrics can remain finite even at a gapless point for $p = q = 0$.
Specifically, we consider the parameter choice,
\begin{equation}
t_2 = 1, \quad t_3 = \frac{1}{5}, 
\quad \gamma_1 = \frac{2}{3}, \quad \gamma_2 = 0.
    \label{param2}
\end{equation}
For this choice of the parameters, the energy gap closes at $t_1 = 13/10$, separating the topologically nontrivial and trivial gapped phases;
see Fig.~\ref{spec-obc-gapped}.

In Fig.~\ref{chi-cancel}\,(a),
we find that
both $p(\theta)$ and $q(\theta)$ vanish at the gap-closing point $\theta=\theta^* = \pi$ [i.e., $p(\theta^*)=q(\theta^*)=0$].
Around this simultaneous zero, 
we expand both $p(\theta)=p|_{\beta=\beta(\theta)}$ and $q(\theta)=q|_{\beta=\beta(\theta)}$ as
\begin{align}
p(\theta)&=
p'(\theta^*)(\theta-\theta^*) + \frac{p''(\theta^*)}{2}(\theta-\theta^*)^2
+ O (\theta-\theta^*)^3, \\
q(\theta)&=
q'(\theta^*)(\theta-\theta^*) + \frac{q''(\theta^*)}{2}(\theta-\theta^*)^2
+ O (\theta-\theta^*)^3,
\end{align}
with
\begin{align}
p'(\theta) \coloneqq \frac{d}{d\theta} \left[p|_{\beta=\beta(\theta)}\right]&,~~ 
q'(\theta) \coloneqq \frac{d}{d\theta} \left[q|_{\beta=\beta(\theta)}\right]; \\
p''(\theta) \coloneqq \frac{d^2}{d\theta^2} \left[p|_{\beta=\beta(\theta)}\right]&,~~
q''(\theta) \coloneqq \frac{d^2}{d\theta^2} \left[q|_{\beta=\beta(\theta)}\right].
\end{align}
Although $q(\theta)$ looks almost quadratic in Fig.~\ref{chi-cancel}\,(a),
this is merely a superficial feature due to a large difference in magnitude between  
$p'(\theta^*)$ and $q'(\theta^*)$.
Recall that in Eqs.~\eqref{eq:gRR_obc} and \eqref{eq:gLR_obc},
the origin of the divergence is the combination
\begin{equation}
\frac{\partial_\theta p(\beta)}{p(\beta)} - \frac{\partial_\theta q(\beta)}{q(\beta)} = \frac{p' (\theta)}{p (\theta)} - \frac{q' (\theta)}{q (\theta)}.
    \label{dpp-dqq}
\end{equation}
Around $\theta = \theta^*$, these terms behave as
\begin{align}
\frac{p' (\theta)}{p (\theta)} &=\frac{1}{\theta-\theta^*} + \frac{p''(\theta^*)}{2 p'(\theta^*)}
+ O (\theta-\theta^*), \\
\frac{q' (\theta)}{q (\theta)} &= \frac{1}{\theta-\theta^*} + \frac{q''(\theta^*)}{2 q'(\theta^*)}
+ O (\theta-\theta^*).
\end{align}
This shows that the leading-order, divergent contributions of order ${1/(\theta-\theta^*)}$ in Eq.~\eqref{dpp-dqq} cancel exactly. 
As a result, Eq.~\eqref{dpp-dqq} converges to a constant:
\begin{equation}
\frac{\partial_\theta p(\beta)}{p(\beta)} - \frac{\partial_\theta q(\beta)}{q(\beta)}
=
\frac{p''(\theta^*)}{2 p'(\theta^*)} -
\frac{q''(\theta^*)}{2 q'(\theta^*)}
+ O (\theta-\theta^*).
\label{dpp-dqq2}
\end{equation}
Accordingly, the asymptotic behaviors of  the quantum metrics are obtained as
\begin{align}
\chi^{\rm RR}(\theta) &\simeq \cfrac{\left| \cfrac{p'(\theta^*)}{q'(\theta^*)} \right|}
{16\left(1+\left| \cfrac{p'(\theta^*)}{q'(\theta^*)} \right| \right)^2}
\left|
\frac{p''(\theta^*)}{p'(\theta^*)} -
\frac{q''(\theta^*)}{q'(\theta^*)}
\right|^2
\nonumber \\
&= {\rm const.},
    \label{chi_RR3}
\end{align}
and
\begin{equation}
\chi^{\rm LR}(\theta) \simeq - \frac{1}{64}
\left(
\frac{p''(\theta^*)}{p'(\theta^*)} - \frac{q''(\theta^*)}{q'(\theta^*)}
\right)^2
={\rm const.}
    \label{chi_LR3}
\end{equation}

Figures~\ref{chi-cancel}\,(b) and (c) show the behavior of
$\chi^{\rm RR}$ and $\chi^{\rm LR}$
for $t_1=13/10$.
The quantum metric $\chi^{\rm RR} (\theta)$ defined solely from right eigenstates exhibits no divergence at $\theta=\pi$.
As approaching the gapless point $\theta\rightarrow\pi$ from either side,
$\chi^{\rm RR} (\theta)$ converges to a finite value consistent with
Eq.~\eqref{chi_RR3}.
Similarly, the biorthogonal quantum metric $\chi^{\rm LR} (\theta)$ does not show the divergence at $\theta=\pi$.

It should be noted that, even in the Hermitian Su-Schrieffer-Heeger model, an analogous mechanism applies, and a gap closing does not necessarily lead to the divergence of the quantum metric.
Instead, the feature unique to the non-Hermitian case is that the off-diagonal components $p$ and $q$ governing the two-band model are no longer constrained by Hermiticity.
Consequently, a generic non-Hermitian gap closing can occur when only one of them vanishes, 
leading to an exceptional point and singular quantum geometry. 
The simultaneous-zero case responsible for the cancellation is rather nongeneric in non-Hermitian systems. 
Thus, non-Hermiticity provides a richer structure for the appearance and disappearance of quantum-metric divergences.

\section{Concluding remarks}
    \label{sec: conclusion}

In this work, we have highlighted various quantum geometric aspects of the non-Hermitian skin effect. 
Primarily, the skin effect manifests itself in the tail/plane-wave part of the wave function, as discussed in Sec.~\ref{sec: Hatano-Nelson}. 
Taking the Hatano-Nelson model as a prototypical example, 
we have shown that the quantum metric $\chi^{\mathrm{RR}}$ in Eq.~\eqref{eq:qmetric_rr_general}
defined solely from right eigenstates captures the skin effect.
The skin effect thus leaves a quantitative and gauge-invariant imprint on quantum geometry.
By contrast, the biorthogonal quantum metric $\chi^{\mathrm{LR}}$ in Eq.~\eqref{eq:qmetric_lr_general} defined from both right and left eigenstates  is unaffected,  
reproducing essentially the result under the periodic boundary conditions.
The difference between right and left eigenstates is also relevant to the Anderson transitions in non-Hermitian systems;
see, for example, Ref.~\cite{Hatano-Nelson-98}.
It is worthwhile to revisit such transitions from the perspective of quantum geometry.

The sensitivity of non-Hermitian systems to boundary conditions 
becomes even more pronounced in models with internal degrees of freedom, 
as illustrated by the non-Hermitian, nonreciprocal Su-Schrieffer-Heeger model in Secs.~\ref{sec: NH-SSH} and \ref{sec: non-Bloch}.
In the Hermitian case, 
the appearance of zero-energy edge states under the open boundary conditions
is attributed to nontrivial topology of the eigenstates under the periodic boundary conditions (i.e., bulk-boundary correspondence). 
In non-Hermitian systems, however, 
both the spectrum and eigenstates are extremely sensitive to boundary conditions, 
and this correspondence superficially breaks down. 
To overcome this difficulty, we have employed the recipe of non-Bloch band theory~\cite{YW-18-SSH, Yokomizo-19}. 
We have demonstrated that the obtained expressions for the quantum metric, as in Eqs.~\eqref{eq:gRR_formula} and \eqref{eq:gLR_formula}
for the periodic boundary conditions, 
and Eqs.~\eqref{eq:gRR_obc} and \eqref{eq:gLR_obc}
for the open boundary conditions,
have a structure compatible with this extreme sensitivity to the boundary conditions. 

We have further focused on the open boundary conditions, 
under which the generalized Brillouin zone plays a crucial role. 
We have shown that nonanalytic points on the generalized Brillouin zone give rise to discontinuities in the quantum metrics. 
Since the skin effect changes the entire structure of the spectrum and eigenstates 
when the boundary conditions are switched from periodic to open, 
both $\chi^{\mathrm{RR}}$ and $\chi^{\mathrm{LR}}$ are sensitive to various singularities on the generalized Brillouin zone.  
As implied in the expressions in Eqs.~\eqref{eq:gRR_obc} and \eqref{eq:gLR_obc},
both $\chi^{\mathrm{RR}}$ and $\chi^{\mathrm{LR}}$
diverge at an isolated zero of $p = p(\beta)$ or $q = q(\beta)$,
which typically arises in the gapless case
studied in Sec.~\ref{sec: gapless}.
On the other hand, at the gap closing where both $p$ and $q$ vanish, 
the divergent contributions to the quantum metrics associated with the two zeros
can cancel, 
resulting in regular behavior of $\chi$'s, as discussed in Sec.~\ref{sec: cancellation}.

Beyond the prototypical examples considered in this work, it is important to study how the present quantum geometric picture extends to more general non-Hermitian systems. 
Generic non-Hermitian systems can host both localized skin modes and delocalized modes within the same spectrum (see, for example, Ref.~\cite{Song-19}), and the resulting quantum geometry should reflect the coexistence of two qualitatively distinct contributions: 
the former yields a finite contribution associated with the skin-localization length, whereas the latter gives a growing contribution characteristic of extended states (see also Sec.~\ref{sec: Hatano-Nelson}).  
Moreover, while this work has been restricted to one-dimensional systems, higher-dimensional non-Hermitian systems exhibit a much richer variety of skin effects~\cite{YSW-18-Chern, Liu-19, Lee-Li-Gong-19, Yoshida-20, Okugawa-20, KSS-20, Fu-21, Zhang-22, Sun-21, Wang-24, Zhang-25}. 
Our geometric characterization should be straightforwardly applicable when open boundary conditions are imposed only along one direction. 
By contrast, the generalization becomes more subtle under open boundary conditions along multiple directions.
It is thus worthwhile to investigate how such higher-dimensional boundary localization phenomena are encoded in quantum geometry and whether the geometric approach facilitates the characterization of higher-dimensional skin effects. 
A geometric characterization of the scale-free skin effect is also worth studying~\cite{Li-20, Yokomizo-21, Wang-23SciPostPhys}.
Furthermore, since quantum geometry remains a meaningful framework also in the presence of disorder and many-body interactions, extending the present approach to disordered and interacting non-Hermitian systems should also be worth investigating.

Finally, the quantum metric plays a significant role in various response and transport phenomena of Hermitian systems~\cite{Yu-25-review, Gao-25-review}.
The non-Hermitian quantum metrics studied in this work may also leave observable signatures in physical phenomena.
Specifically, it is significant to find natural physical observables that detect nonanalytic points of the generalized Brillouin zone, which are a feature unique to non-Hermitian systems, so that such singularities become experimentally accessible. 
More broadly, it merits further investigation to elucidate which observables provide the most direct probes of quantum geometric effects in non-Hermitian systems.

\medskip
\begingroup
\renewcommand{\addcontentsline}[3]{}
\begin{acknowledgments}
K.I. is indebted to Naoto Nagaosa for introducing him to quantum geometry and for encouraging the exploration of the non-Hermitian case.
We thank RIKEN iTHEMS, where this work was partially completed during the workshop on ``de Sitter Holography Meets Non-Hermitian Quantum Matter".
K.I. is supported by JSPS KAKENHI Grant No.~JP24K00545.
K.K. is supported by JSPS KAKENHI Grant No.~JP24H00945, No.~JP26H02015, No.~JP26K06970, and No.~JP26K17046, and JST FOREST Program Grant No.~JPMJFR256P.
\end{acknowledgments}
\endgroup



\let\oldaddcontentsline\addcontentsline
\renewcommand{\addcontentsline}[3]{}
\bibliography{ref.bib}
\let\addcontentsline\oldaddcontentsline

\end{document}